\documentclass[twocolumn]{aastex62}
\usepackage{rotating}
\usepackage{float}

\newcommand{\as}{^{\prime \prime}}
\newcommand{\kms}{\rm km~s^{-1}}
\newcommand{\nthp}{\rm N_{\rm 2}H^{\rm +}}
\newcommand{\hcop}{\rm HCO^{\rm +}}
\newcommand{\ceo}{\rm C^{\rm 18}O}
\newcommand{\tco}{\rm ^{\rm 13}CO}
\newcommand{\nht}{N_{\rm H_{2}}}
\newcommand{\mlin}{M_{\rm line}}

\newcommand{\vpeak}{V_{\rm peak}}
\newcommand{\tex}{T_{\rm ex}}
\newcommand{\lfil}{L_{\rm fil}}

\accepted{for publication in the ApJ}

\shorttitle{TRAO FUNS I. L1478 in the California MC}
\shortauthors{Chung et al.}

\begin{document}

\title{TRAO Survey of Nearby Filamentary Molecular clouds, the Universal Nursery of Stars (TRAO FUNS) I. \\ Dynamics and Chemistry of L1478 in the California Molecular Cloud} 
  

\author{Eun Jung Chung}
\affil{Korea Astronomy and Space Science Institute, 776 Daedeokdae-ro, Yuseong-gu, Daejeon 34055, Republic of Korea}

\author{Chang Won Lee}
\affil{Korea Astronomy and Space Science Institute, 776 Daedeokdae-ro, Yuseong-gu, Daejeon 34055, Republic of Korea}

\author{Shinyoung Kim}
\affil{Korea Astronomy and Space Science Institute, 776 Daedeokdae-ro, Yuseong-gu, Daejeon 34055, Republic of Korea}

\author{Gwanjeong Kim}
\affiliation{Nobeyama Radio Observatory, National Astronomical Observatory of Japan, Nagano 384-1305, Japan}

\author{Paola Caselli}
\affiliation{Max-Planck-Institut f$\ddot{u}$r Extraterrestrische Physik, D-85748 Garching, Germany}

\author{Mario Tafalla}
\affiliation{Observatorio Astron$\acute{o}$mico Nacional (IGN), Alfonso XII 3, 28014 Madrid, Spain}

\author{Philip C. Myers}
\affiliation{Harvard-Smithsonian Center for Astrophysics, 60 Garden Street, Cambridge, MA 02138, USA}

\author{Archana Soam}
\affil{SOFIA Science Centre, USRA, NASA Ames Research Centre, MS N232, Moffett Field, CA 94035, USA}

\author{Tie Liu}
\affil{Korea Astronomy and Space Science Institute, 776 Daedeokdae-ro, Yuseong-gu, Daejeon 34055, Republic of Korea}

\author{Maheswar Gopinathan}
\affil{Indian Institute of Astrophysics, Koramangala, Bangalore 560034, India}

\author{Miryang Kim}
\affil{Korea Astronomy and Space Science Institute, 776 Daedeokdae-ro, Yuseong-gu, Daejeon 34055, Republic of Korea}

\author{Kyoung Hee Kim}
\affiliation{Department of Earth Science Education, Kongju National University, 56 Gongjudaehak-ro, Gongju-si, Chungcheongnam-do 32588, Korea}

\author{Woojin Kwon}
\affil{Korea Astronomy and Space Science Institute, 776 Daedeokdae-ro, Yuseong-gu, Daejeon 34055, Republic of Korea}

\author{Hyunwoo Kang}
\affil{Korea Astronomy and Space Science Institute, 776 Daedeokdae-ro, Yuseong-gu, Daejeon 34055, Republic of Korea}

\author{Changhoon Lee}
\affil{Korea Astronomy and Space Science Institute, 776 Daedeokdae-ro, Yuseong-gu, Daejeon 34055, Republic of Korea}

\begin{abstract}
$^{\prime \prime}$TRAO FUNS$^{\prime \prime}$ is a project to survey Gould Belt's clouds in molecular lines. This paper presents its first results on the central region of the California molecular cloud, L1478. We performed On-The-Fly mapping observations using the Taedeok Radio Astronomy Observatory (TRAO) 14m single dish telescope equipped with a 16 multi-beam array covering $\sim$1.0 square degree area of this region using $\ceo (1-0)$ mainly tracing low density cloud and about 460 square arcminute area using $\nthp (1-0)$ mainly tracing dense cores. $\rm CS (2-1)$ and $\rm SO (3_{2}-2_{1})$ were also used simultaneously to map $\sim$440 square arcminute area of this region. We identified 10 filaments by applying the dendrogram technique to the $\ceo$ data-cube and 8 dense $\nthp$ cores by using {\sc FellWalker}. Basic physical properties of filaments such as mass, length, width, velocity field, and velocity dispersion are derived. It is found that L1478 consists of several filaments with slightly different velocities. Especially the filaments which are supercritical  are found to contain dense cores detected in $\nthp$. Comparison of non-thermal velocity dispersions derived from $\ceo$ and $\nthp$ for the filaments and dense cores indicates that some of dense cores share similar kinematics with those of the surrounding filaments while several dense cores have different kinematics with those of their filaments. This suggests that the formation mechanism of dense cores and filaments can be different in individual filaments depending on their morphologies and environments. \\
\end{abstract}


\section{Introduction} \label{sec:intro}

How stars form in molecular clouds is one of the key questions in astronomy. In general stars are known to form by a gravitational contraction in dense cores which are made in less dense molecular clouds by their hierarchical fragmentation. Recent high resolution observations mainly done by $Spitzer$ Space Telescope\footnote{\url{https://www.cfa.harvard.edu/gouldbelt/}} and $Herschel$ Space Observatory\footnote{\url{http://www.herschel.fr/cea/gouldbelt/en/}} reveal that molecular clouds are filamentary and such a structure is ubiquitous over various star-forming environments from active star-forming molecular clouds like the Orion molecular complex to non-star-forming molecular clouds such as the Polaris flare \citep[e.g.,][]{andre2010,hacar2018}. 

During last decade, several studies have been done and progresses are made in understanding the physical properties of filaments and dense cores. One of the most interesting findings about filaments is that filaments have a characteristic width of 0.1~pc which is comparable to the typical size of dense cores \citep[e.g.,][]{andre2010,arzoumanian2011,palmeirim2013,federrath2016,arzoumanian2019}. Most prestellar cores are found on the dense, supercritical filaments where the mass per unit length is larger than the critical value of isothermal cylinders \citep[e.g.,][]{andre2010,konyves2015,marsh2016}. 
Many young stellar groups in the nearby molecular clouds are found to be well associated with Hub-filament structure that consists of a hub which is a central body with relatively higher column density ($>10^{22}~\rm cm^{-2}$) and filaments radiated from the hub which has lower column density \citep[e.g.,][and references therein]{myers2009}. Besides, it is observed velocity gradients along filaments and the gas flow along filaments is responsible for the formation of the star cluster \citep[e.g.,][]{kirk2013,peretto2014,imara2017,baug2018,yuan2018}. Hence, it seems clear that filaments can play a crucial role in the formation of cores and stars. 

These results raise important questions about such as 1) how filaments and dense cores form in large molecular clouds? 2) are filaments an intermediate stage of star formation, i.e., from large clouds to dense cores? These can be answered by probing the kinematics and chemistry of filaments and dense cores with systematic molecular line observations toward various molecular clouds. 

We have been performing such observations for filament clouds using Taedeok Radio Astronomy Observatory (TRAO)\footnote{\url{http://radio.kasi.re.kr/trao/main_trao.php}} 14m antenna with a project $^{\prime \prime}$TRAO FUNS$^{\prime \prime}$ which is an acronym for $^{\prime \prime}$the TRAO survey of Filaments, the Universal Nursery of Stars$^{\prime \prime}$. This project is to make a systematic survey for 10 Gould Belt's clouds with several molecular lines in various environments, aiming to obtain 1) the velocity structure of filaments and dense cores for the study of their formation, 2) radial accretion or inward motions toward dense cores from their surrounding filaments, and 3) chemical differentiation of filaments and their dense cores. For these goals, six molecular lines of $\ceo ~(1-0)$, $\tco ~(1-0)$, $\nthp ~(1-0)$, $\hcop ~(1-0)$, $\rm SO ~(3_{2}-2_{1})$, and $\rm CS ~(2-1)$ are selected as tracers of the kinematics and chemical evolution for the filaments and dense core. More details on the TRAO FUNS will be introduced in other paper by Lee et al. (in preparation).

In this paper, we present the first results of the TRAO FUNS survey toward L1478 in the California molecular cloud. The California molecular cloud (CMC, hereafter) which is also called the Auriga-California molecular cloud has been recently recognized as a massive giant molecular cloud by \citet{lada2009}. They used infrared extinction map from the Two Micron All Sky Survey \citep[2MASS;][]{Kleinmann1994} and CO maps from the Galactic plane survey \citep{dame2001}, and found continuous distribution of the molecular cloud in velocity and space as a single molecular cloud at the same distance. CMC is located at a distance of 450$\pm$23 pc, and it is comparable in size ($\sim$80 pc) and mass ($\sim 10^{5} ~ M_{\odot}$) to the Orion giant molecular cloud (OMC). However, the number of young stellar objects (YSOs) in CMC (149) is 15-20 times  smaller than that of OMC \citep[3330,][]{broekhoven-fiene2014}. \citet{harvey2013} investigated the young stellar objects and dense gas of CMC, finding 60 compact sources at 70/160 $\mu m$ and 11 cold, compact sources at 1.1 mm. Recently, \citet{broekhoven-fiene2018} identified 59 candidate protostars, and found that 24 among them are associated with YSOs in the catalogs of $Spitzer$ and $Herschel/$PACS. They suggested that CMC is significantly less efficient in star formation than Orion A. There are recent observations of molecular lines for a part of CMC by \citet{imara2017}. They investigated the relationship of filaments and dense cores in the CMC region ($\sim 0.5^{\circ} \times 0.5^{\circ}$ area) with dust continuum data ($Herschel$) and $^{12}$CO and $^{13}$CO (2-1) molecular line data obtained from the Heinrich Hertz Submillimeter Telescope, finding  that filaments in the west region of L1478 are velocity-coherent and gravitationally supercritical. 

We have investigated filaments and dense cores of CMC, L1478 ($\sim 2^{\circ} \times 0.5^{\circ}$), which provides a good laboratory to investigate the formation of filament and dense cores. Various morphologies of filaments such as a long network of filaments and a hub-filaments structure can be found (see Figure 1
). Besides, the star forming property of L1478 is relatively modest and seems to be a low mass star forming region, in a sense that only three YSOs are found, and thus can be a good comparison to the OMC \citep{broekhoven-fiene2018}. Our survey provides the first mapping observations in various molecule lines toward the central region of the CMC.

This paper is organized as follows. In section~2
, observation and data reduction are explained. Identifications of filament and dense core and their basic physical properties are presented in section~3
. In section~4
, we discuss about velocity structure of filaments, gravitational instability, and the formation mechanisms of filaments and dense cores. Section~5 
summarizes our results. \\

\section{Observations and Data Reduction} \label{sec:obs}

\subsection{Observations}

We have carried out On-The-Fly (OTF) mapping observations toward L1478 in CMC with TRAO 14-m telescope from January to May in 2017 and from December 2017 to February 2018. The equipped frontend is SEcond QUabbin Optical Image Array (SEQUOIA-TRAO), which consists of a 4$\times$4 array receiver and its spatial separation is 89$^{\prime \prime}$. The backend, FFT spectrometer, is 4096$\times$2 channels at 15~kHz resolution ($\sim 0.04~\kms$ at 110~GHz) which covers a total bandwidth of 62.5~MHz corresponding to $\sim 170~\kms$ at 110~GHz. TRAO systems allow simultaneous observations of 2 molecular lines between the frequency range of 85 and 100~GHz or 100 and 115~GHz. The beam efficiency at 90~GHz is 0.48 and at 110~GHz is 0.46.  

\begin{figure*}
\plotone{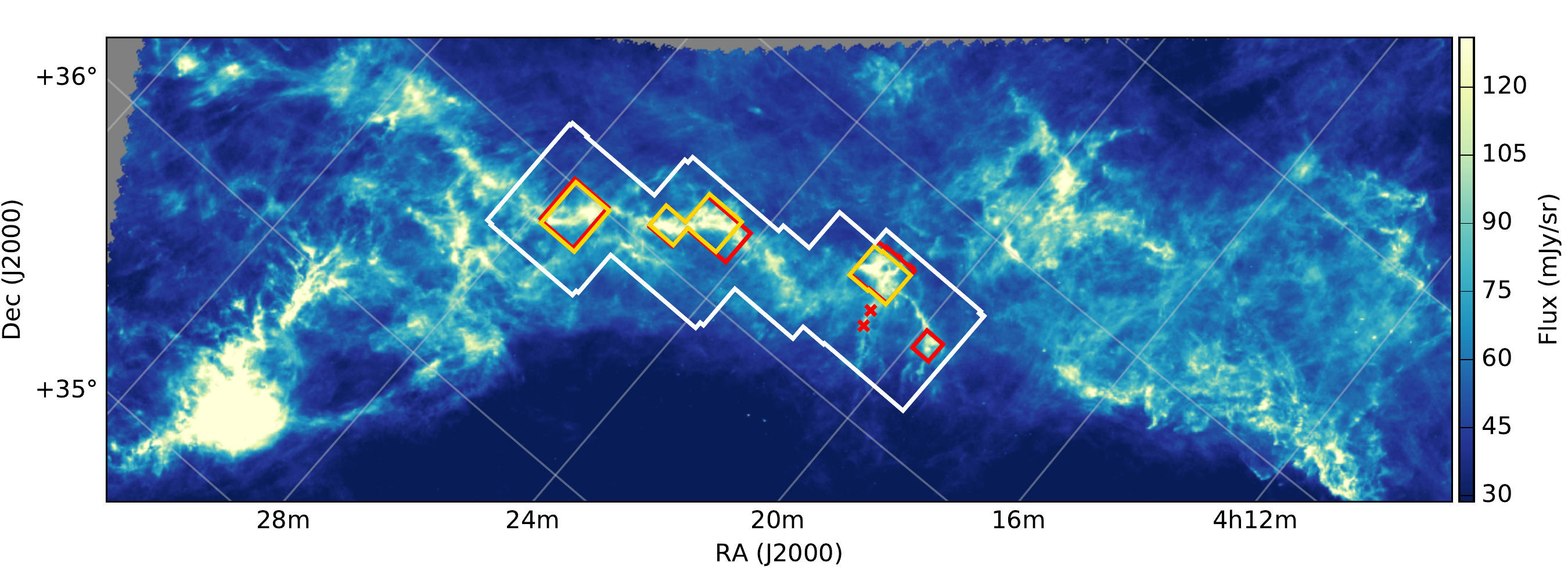}
\caption{$Herschel$ 250$\mu$m image of the Auriga-California region. The area covered in this study, L1478, is indicated by white ($\tco$ and $\ceo ~1-0$), red ($\nthp$ and $\hcop ~1-0$), and yellow (SO $3_{2}-2_{1}$ and CS $2-1$) boxes. The red crosses show the region where position-switching mode observations of $\nthp$ and $\hcop$ molecular lines have been done. \label{fig:acm}}
\end{figure*}

To investigate the physical properties of filaments and dense cores, six molecular lines are observed. $\ceo$ and $\nthp(1-0)$ molecular lines are chosen as a tracer of relatively less dense material for the filaments and a dense gas tracer for dense cores, respectively. $\tco(1-0)$ which can reveal the large scale bulk motion is simultaneously obtained with $\ceo$. With $\nthp$ observation, $\hcop (1-0)$ is concurrently observed. To probe the chemical evolution of dense cores, SO ($3_{2}-2_{1}$) is selected and CS ($2-1$) observed at the same time. 
SO(32-21) is known as one of the most sensitive molecules to the depletion and hence can be used as a tracer of very young dense cores \citep{tafalla2006}. CS(2-1) line was chosen to be useful to study infall motions in the pre-stellar cores \citep{lee2001}. In this study, four molecular lines of $\ceo$, $\nthp$, CS, and SO are mainly used in analyses. 

\floattable 
\begin{deluxetable*}{lccCccc} 
\tablecaption{Observations \label{tab:lines}} 
\tablecolumns{5} \tablewidth{0pt} 
\tablehead{ 
\colhead{Molecule} & 
\colhead{$\nu_{ref}$\tablenotemark{a}} & 
\colhead{$\theta_{\rm FWHM}$\tablenotemark{b}} & 
\colhead{Area\tablenotemark{c}} & 
\colhead{$\theta_{\rm pixel}$\tablenotemark{d}} & 
\colhead{$\delta v$\tablenotemark{e}} & 
\colhead{$rms$\tablenotemark{f}} \\
\colhead{} & 
\colhead{(GHz)} & 
\colhead{($\as$)} & 
\colhead{(sq. arcmin)} & 
\colhead{($\as$)} & 
\colhead{($\kms$)} & 
\colhead{(K[T$_{\rm A}^{\ast}$])} } 
\startdata
$\ceo ~(1-0)$ & 109.782160 & 47 & $\sim 3700$ & 44 & 0.1 & 0.092  \\
$\tco ~(1-0)$ & 110.201353 & 47 & $\sim 3700$ & 44 & 0.1 & 0.097  \\
$\nthp ~(1-0)$ & 93.173764 & 56 & $\sim 460$ & 22 & 0.06 & 0.063  \\
$\hcop ~(1-0)$ & 89.188525 & 58 & $\sim 460$ & 22 & 0.06 & 0.062  \\
SO $(3_{2}-2_{1})$ & 99.299870 & 52 & $\sim 440$ & 22 & 0.06 & 0.094  \\
CS $(2-1)$ & 97.980953 & 52 & $\sim 440$ & 22 & 0.06 & 0.097  \\
\enddata 
\tablenotemark{}{} \\
\tablenotemark{a}{~Rest frequency of each molecular line is taken from The Cologne Database for Molecular Spectroscopy \citep[CDMS:][https://cdms.ph1.uni-koeln.de/ cdms/portal/]{muller2001}. } \\
\tablenotemark{b}{~Full width half maximum of the beam} \\
\tablenotemark{c}{~Total observed area} \\
\tablenotemark{d,e}{~The pixel size and channel width of the final datacube} \\
\tablenotemark{f}{~Noise level in T$_{\rm A}^{\ast}$ of the final datacube} \\
\end{deluxetable*}

We divided our target area of $\ceo$ (and $\tco$ simultaneously) into five regions which have a box shape referred as `tiles' hereafter, shown in Figure~1
, and carried out OTF mapping observations. Each tile has a size from 12$^{\prime} \times 12^{\prime}$ to 32$^{\prime} \times 32^{\prime}$. The scanning rate was 55$\as$ per second and the integration time is 0.2 second. Scan step of 0.25~HPBW (44$\as$) along the scan direction and 0.75~HPBW separation between the rows are applied for $\ceo$ and $\tco$. We carried out OTF mapping observations of $\nthp$ and $\hcop$  towards the regions only where $\ceo$ is strongly detected. Five tiles were observed with a scan step of 0.25~HPBW (44$\as$) along the scan direction and a 0.25~HPBW separation between the rows. Simultaneous observation of SO and CS lines are performed with the same OTF mapping parameters as the observations of $\ceo$ and $\tco$, but for only four tiles. We made maps alternatively along RA and Dec directions. In Figure~1
, The observed area of the six molecular lines are presented with the $Herschel$ 250~$\mu$m continuum image. Toward two points where $\ceo$ is strongly detected but not carried out OTF observation of $\nthp$ and $\hcop$, additional position-switching (PS) observations of $\nthp$ (and $\hcop$) have been carried out due to the lack of observing time (denoted with red crosses in Figure~1
. The $rms$ noise level of PS observations is about 0.06~K in the unit of antenna temperature for both lines. \\

\subsection{Data Reduction}

The raw OTF data for each map of each tile is read and produced into a map with jinc-gaussian function after baseline fitting (with 1st order) in {\sc Otftool}. We give the resulted cell size of 22$^{\prime \prime}$ and apply noise-weighting. Further reduction and examinations are done with the {\sc Class}\footnote{\url{http://www.iram.fr/IRAMFR/GILDAS}} package. Since the baseline is not good in both ends of the band and the total velocity range of the spectra ($\sim \rm 170~\kms$) is much larger than that of emissions in our object (less than 20~$\kms$), baseline fittings are done in two steps. Firstly, both ends of the spectra were cut off so that the velocity range becomes 120~$\kms$ and a baseline was subtracted with the second order polynomial. After that, the spectra are resampled with channel width of 0.06~$\kms$, both ends were cut off again resulting in a spectral velocity range of 60~$\kms$, and baseline subtraction was done again with first order polynomial. After this step, all channel maps are examined and maps having high noise level or showing some spatial gradient of noise level due to the change of system temperature were excluded. Finally, the maps are merged into a final fits cube with 44$^{\prime \prime}$ cell size and 0.1~$\kms$ velocity channel width for $\ceo$ and $\tco$, and 22$^{\prime \prime}$ cell size and 0.06~$\kms$ velocity channel width for the other molecular lines. The basic observational information  is given in Table~1
. \\

\begin{figure*}
\plotone{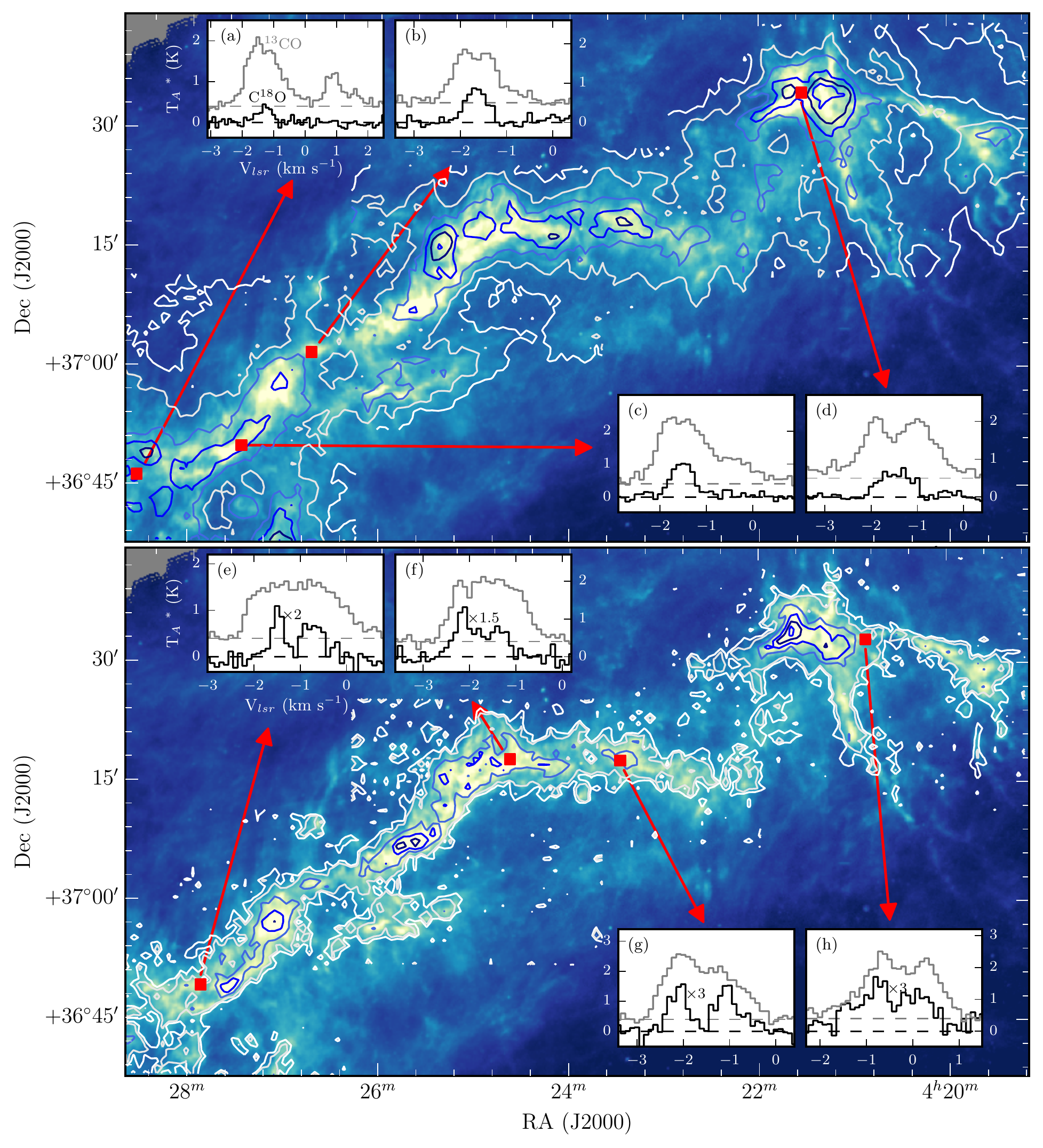}
\caption{Integrated intensity maps of $\tco$(1-0) (top) and $\ceo$(1-0) (bottom) toward L1478. Background color images are $Herschel$ 250 $\mu$m data and the CO intensity maps are shown in contours.  The line intensity maps are integrated over a velocity range of $-$3.8 to 2.3 km~s$^{-1}$ for $\tco$, and $-$3.2 to 0.6~$\kms$ for $\ceo$. The contour levels of $\tco$ are 7$n \times \sigma_{\rm rms}$ ($n=1, 2, \cdots, 6$) and those of $\ceo$ are 2 and 3$n \times \sigma_{\rm rms}$ ($n=1,2, \cdots, 6$). Two spectra of $\tco$ (gray) and $\ceo$ (black) are given in (a)-(h) inset windows at the selected positions.  The spectra in (a)-(d) windows are given to illustrate  that $\ceo$ traces well the velocity field of filament material while $\tco$ more or less self-absorbed due to its larger optical depth compared with $\ceo$. The spectra in (e)-(h) are shown to indicate that $\ceo$ profiles have two components and thus some of filaments can have the multiple velocity structure to the line-of-sight.  \label{fig:13c18om0}}
\end{figure*}

\section{Filaments and Dense Cores} \label{sec:fila} 

\subsection{Filament Identification} \label{sec:filid}

Figure~2
 shows the integrated intensity maps of $\tco$ (top) and $\ceo$ (bottom) on $Herschel$ 250 $\mu$m image. The distributions of molecules seem to well match that of dust continuum. The long filamentary structure from the southeast to the center of the observed area is well revealed by $\ceo$ as well as $\tco$. It spreads out to $\sim$1 degree in the sky. The other noticeable feature is the hub-filament structure in the northwest that is referred as Cal-X due to its X shape by \citet{imara2017}. The hub radiates four filaments to the east, south, west, and north. 

$\tco$ is detected at the outer edge of the observed regions, while $\ceo$ is well matched to the filamentary dust emission. Most of $\tco$ spectra show multiple velocity components. In the top panel, $\tco$ emission is detected at $\sim 1 ~\kms$ and $\sim -1.5 ~ \kms$, while $\ceo$ is only detected at $\sim -1.5~ \kms$ with a single Gaussian shape indicating that the double peak component at $\sim -1.5~ \kms$ of $\tco$ is a self-absorption feature. Therefore, $\tco$ spectra are thought to be self-absorbed toward some dense regions shown in the spectra (a) to (d), because it has relatively larger optical depth than that of $\ceo$. However, both $\tco$ and $\ceo$ spectra at several positions drawn in the bottom panel of Figure~2
 shows double peaked features, implying that the filaments at these places may consist of multiple components to the line-of-sight.

To identify filaments with multiple velocity structure, we used $astrodendro$\footnote{\url{https://dendrograms.readthedocs.io}} Python package and applied it to the $\ceo$ data cube. A dendrogram is a tree diagram which shows how and where the structures merge and its algorithm identifies the hierarchical structure of 2- and 3-dimensional datasets. Structures start from local maxima, their volumes get bigger merging with the surroundings with lower flux densities, and stop when they meet neighboring structures \citep{rosolowsky2008}. Details of filament identification using the dendrogram technique are given in the Appendix.

\begin{figure*}[ht!]
\gridline{\fig{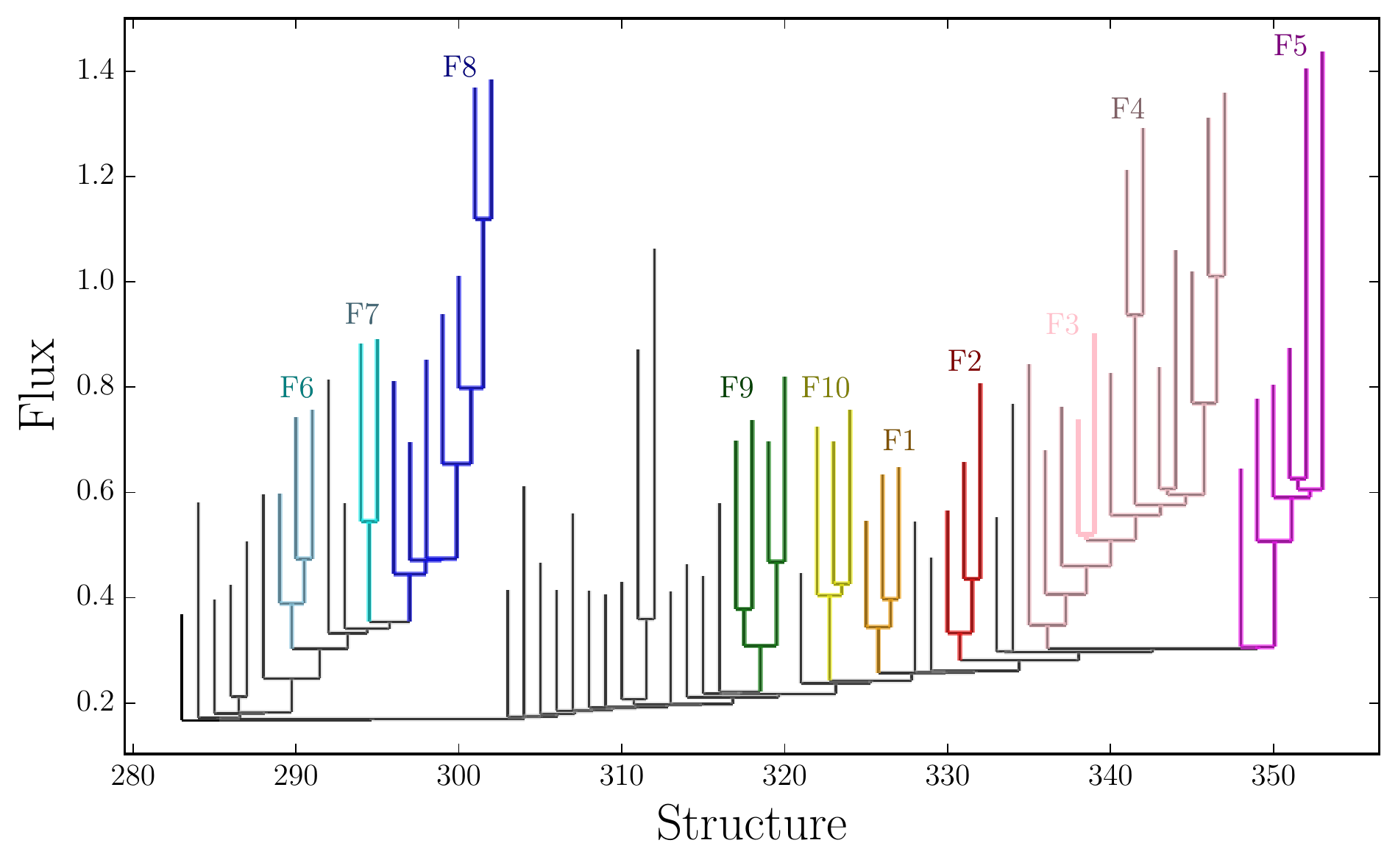}{.83\textwidth}{}} 
\gridline{\fig{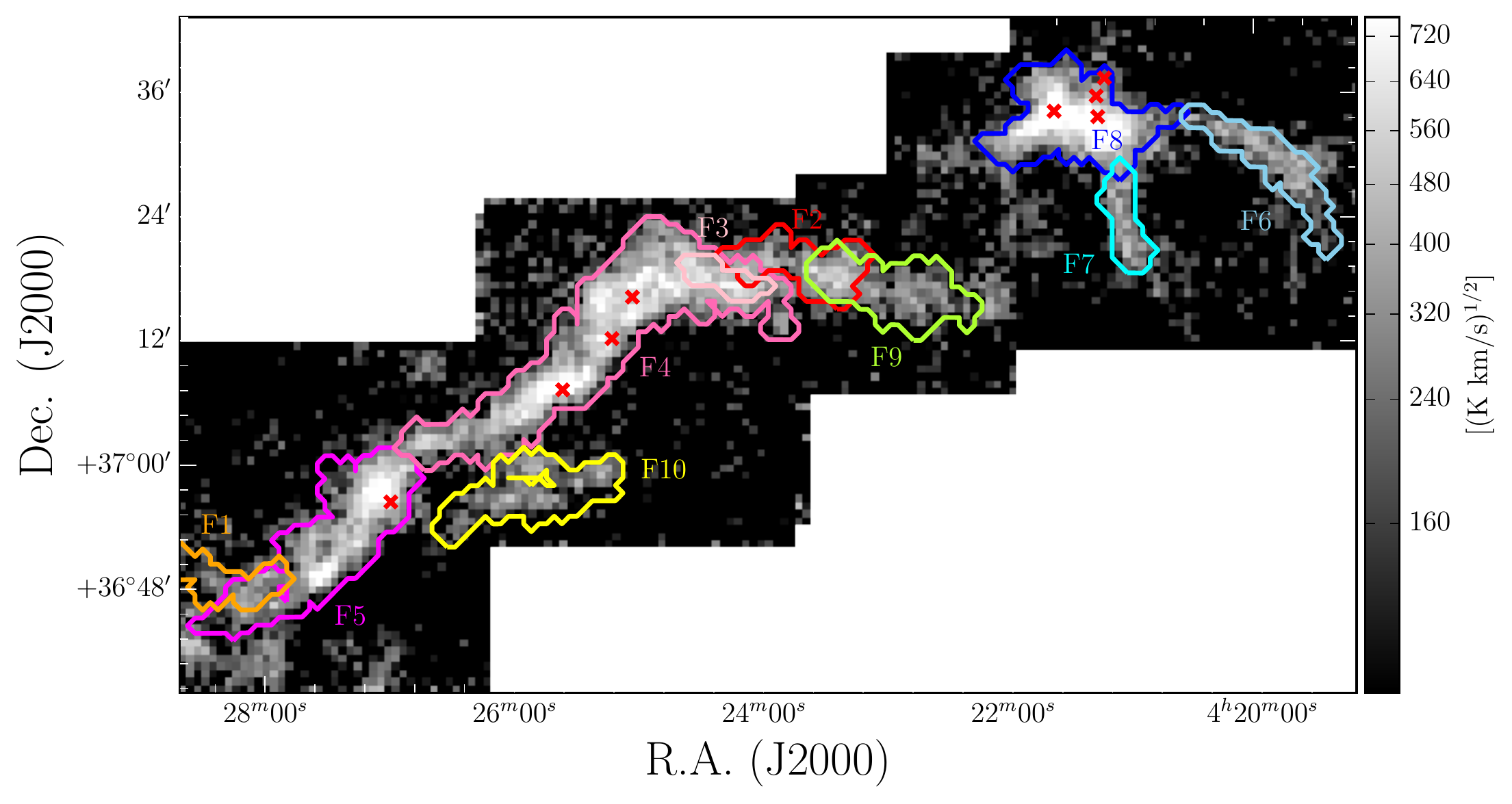}{.9\textwidth}{}}
\caption{\textbf{Top :} Branches and leaves for L1478 identified by dendrogram analysis. The x-axis indicates an identification number of a structure in L1478 and y-axis denotes the intensity of each structure in $\rm K[T_{A}^{\ast}]$. \textbf{Bottom :} Leaves in the dendrogram overlaid on the C$^{\rm 18}$O(1-0) integrated intensity map (the grayscale image). Each identified leaves in the top panel are drawn in this panel as filament, named as F1 to F10. Leaves found in the observing boundary are usually noisy structures falsely identified and thus excluded in further analyses. The red crosses present the locations of dense cores identified with $\nthp$ data (Section~3.2
). \label{fig:dendrogram}}
\end{figure*}

Figure~3
 shows the results of this dendrogram analysis. Resulted branches and leaves are shown in the dendrogram tree (top) and its spatial distribution in the filament clouds is presented in the bottom panel on the $\ceo$ moment 0 image. Each identified leave is indicated with color-coded base on their numbers, i.e., the same color in the tree and map. There are five false leaves found due to noisy observation near the boundary of the map. We excluded them and used the other 10 filaments in the analysis. With the given mask in dendrogram, we extract datacubes of each filaments and use them for further analyses, i.e., central velocity and velocity dispersion. It is noticeable here that F6 (filament 6), F7, and F8 are identified as an independent leaf but they have a single stem, and this is the same for filaments from F1 to F5. Only F9 and F10 have their own stems. We will discuss about these filaments later. \\

\subsection{Dense Core Identification} \label{sec:iddcore}

The $\nthp (1-0)$ molecular line, which is usually optically thin, is an appropriate tracer of dense cores in nearby star forming regions \citep[e.g.,][]{caselli1995,sanhueza2012}. To probe the relationship of dense cores and filaments, we made five OTF tiles of $\nthp$ toward the regions where $\ceo$ emission is strongly detected. Two tiles are in F4, one in F5, one in F6 and one in F8 (see Figures~1
 and 3
 ). $\nthp$ emission is detected in all the observed tiles except in F6. We also carried out PS observations toward two positions in F7 (marked with red crosses in Figure~1
 ), but $\nthp$ was not detected at the $rms$ level of 0.06~K[$\rm T_{\rm A}^{\ast}$]. 

\begin{figure*}
\plotone{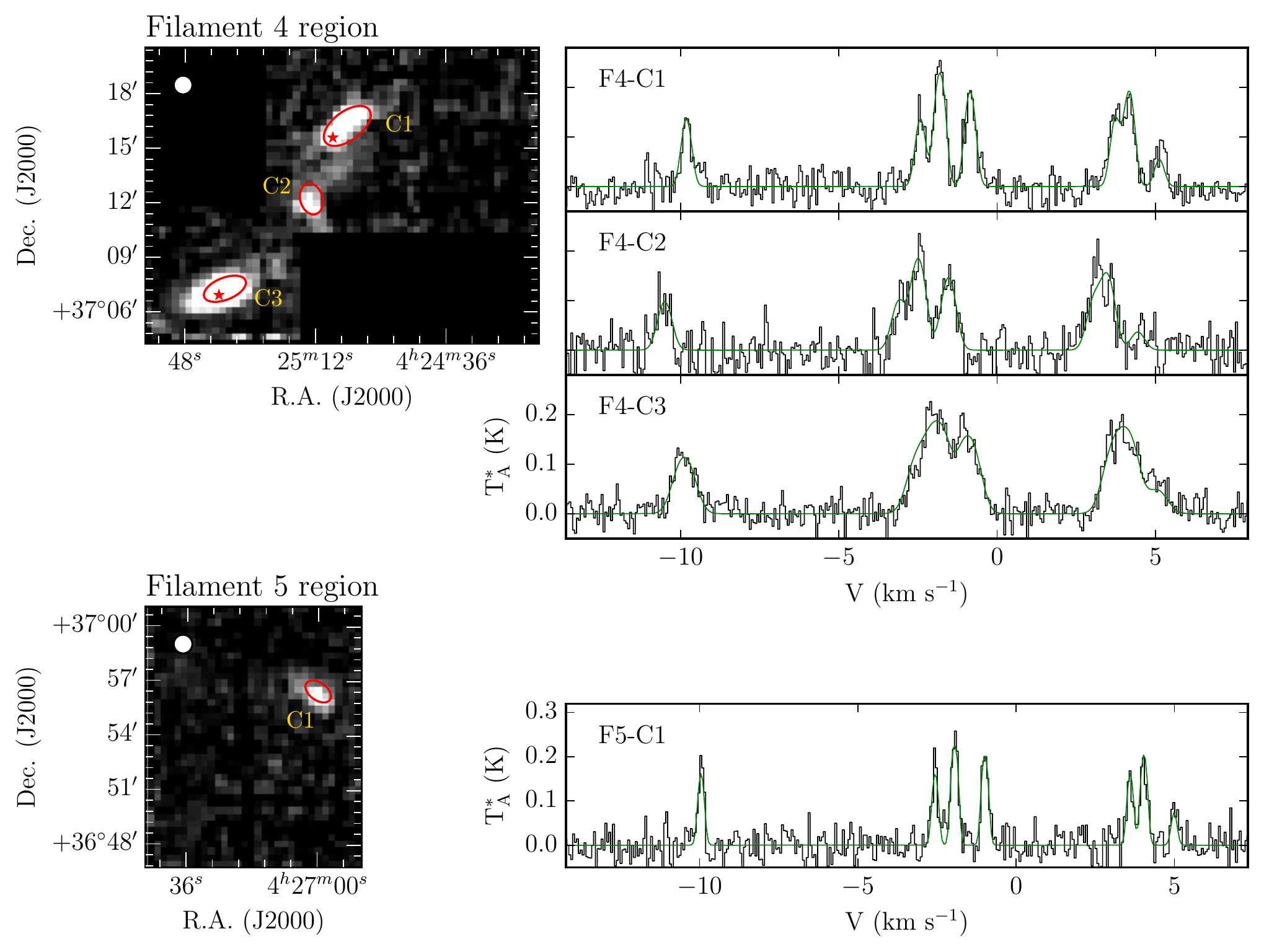}
\caption{\textbf{Left :} Identified dense cores using \textsc{FellWalker} is presented with red ellipses on the $\nthp$ integrated intensity maps (moment 0 maps) of the Filament 4 and Filament 5. The 56$\as$ FWHM beam at 93.176~GHz is shown by the white circle. The red stars represent positions of YSOs reported by \citet{broekhoven-fiene2018}. \textbf{Right :} $\nthp$ spectra of dense cores are presented with hyperfine fitting results (green lines). \label{fig:corespec1}}
\end{figure*}
\begin{figure*}
\plotone{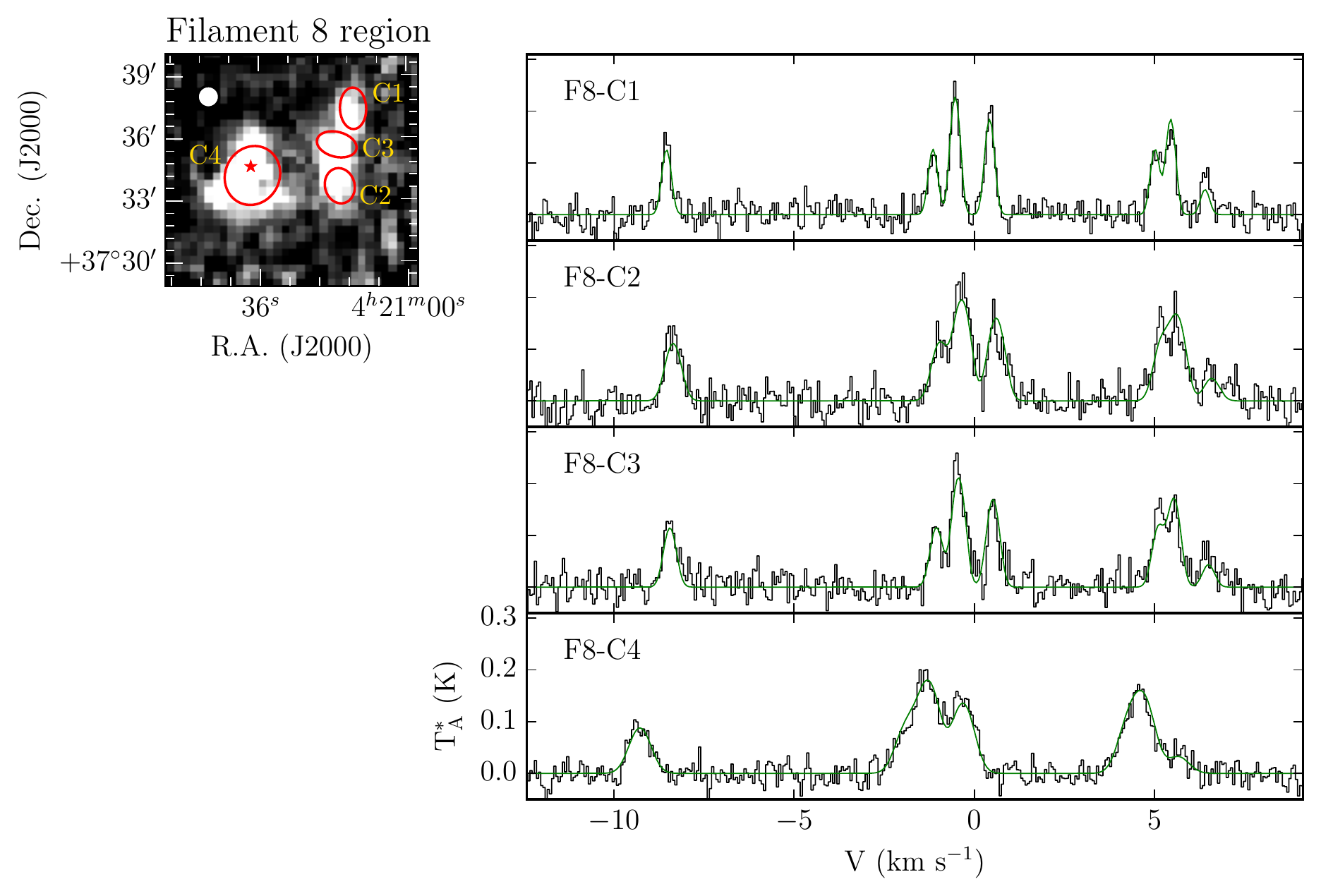}
\caption{Same as Figure~4
 for Filament 8.  \label{fig:corespec2}}
\end{figure*}

We applied the \textsc{FellWalker} source extraction algorithm \citep{berry2015} to the $\nthp$ integrated intensity image to find dense cores. In running this algorithm only pixels whose intensities are higher than 3$\sigma$ were considered. The required minimum number of pixels that a core should include to be identified as a real core is seven, i.e., it should be larger than one beam size of 56$^{\as}$. In case there are neighboring peaks, if the difference between the peak value and the minimum value (dip value) is larger than 1$\sigma_{\rm rms}$, it is considered as an independent core. With these criteria, we obtained eight cores, three in F4, one in F5, and four in F8.  

The masses of the identified dense core are derived with integrated intensity of $\nthp$. We calculated total column density of $\nthp$ following Equation (A4) of \citet{caselli2002ii} and converted it into the corresponding H$_2$ column density with an average abundance of $\nthp$ of $\sim6.8(\pm 4.8) \times 10^{-10}$ \citep{johnstone2010,lee2011}. The uncertainties in the $\nthp$ intensities are typically less then 30\%, while the uncertainties of excitation temperature and the conversion factor between the column densities of $\nthp$ and H$_{2}$ are quite large but less than a factor of 2 \citep{johnstone2010}. Hence, the uncertainties of dense core masses are less than a factor of 2. The virial masses ($M_{\rm vir}$) are derived with the equation of $M_{\rm vir} = k ~ R~ \sigma_{\rm tot}^{2} / G$ where $R$ and $\sigma$ are the radius and total velocity dispersion of the core, respectively \citep{maclaren1988}. For simplicity, we used $k=1$ assuming a density profile of $\rho \propto R^{-2} $ where $R$ is the core radius. The dense cores identified show virial parameters, $\alpha = M_{\rm vir}/M$, between $\sim$0.8 and 2.9. Five dense cores which have $\alpha < 2$ are likely close to a gravitationally bound status. However, we should be careful to interpret the results because the resulted masses have large uncertainties. The information of identified dense cores and calculated masses are listed in Table~2
 and the positions and the spectra are presented in Figure~4
  and 5
  . \\

\begin{deluxetable*}{lccccccccccccl}
\tablecaption{Information of the Identified Dense Cores \label{tab:ppcore}}
\tablecolumns{13}
\tablewidth{0pt}
\tablehead{
\colhead{Core ID} &
\multicolumn{2}{c}{Position} &
\colhead{} &
\multicolumn{2}{c}{Size} & 
\colhead{PA} &
\colhead{$\vpeak (\nthp)$} &
\colhead{$\Delta V( \nthp ) $} &
\colhead{$M_{\rm obs}$} &
\colhead{${\alpha_{\rm vir}}^{\ast}$} &
\colhead{IR Class\tablenotemark{$\dagger$}} \\
\cline{2-3} \cline{5-6} 
\colhead{} &
\colhead{RA} &
\colhead{Dec} &
\colhead{} &
\colhead{Major} & 
\colhead{Minor} &
\colhead{} &
\colhead{} & 
\colhead{} &
\colhead{} \\ 
\colhead{} &
\colhead{(hh:mm:ss)} &
\colhead{(dd:mm:ss)} & 
\colhead{} &
\colhead{(pc)} & 
\colhead{(pc)} & 
\colhead{(deg.)} &
\colhead{($\kms$)} &
\colhead{($\kms$)} &
\colhead{($M_{\odot}$)} &
\colhead{} &
\colhead{} }
\startdata
\multicolumn{3}{l}{Filament 4 region} & & & & & & & & \\
\cline{1-2}
F4-C1 & 04:25:03.21 & +37:16:14.34 & & 0.39 & 0.21 & 126 & -1.78 & 0.35 & 1.42 
& 1.35 & Class 0 \\
F4-C2 & 04:25:13.10 & +37:12:13.32 & & 0.23 & 0.16 & 15 & -2.47 & 0.51 & 0.62 
& 2.88 \\
F4-C3 & 04:25:36.96 & +37:07:16.54 & & 0.32 & 0.16 & 111 & -1.85 & 0.72 & 1.87 
& 1.86 & Class 0/I \\
\multicolumn{3}{l}{Filament 5 region} & & & & & & & & \\
\cline{1-2}
F5-C1 & 04:26:59.95 & +36:56:26.66 & & 0.21 & 0.13 & 54 & -1.92 & 0.21 & 0.33 
& 2.45 \\
\multicolumn{3}{l}{Filament 8 region} & & & & & & & & \\
\cline{1-2}
F8-C1 & 04:21:13.02 & +37:37:23.74 & & 0.26 & 0.17 & 1 & -0.50 & 0.29 & 1.55 
& 0.77 \\
F8-C2 & 04:21:16.44 & +37:33:39.67 & & 0.23 & 0.19 & 14 & -0.32 & 0.49 & 1.80 
& 1.04 \\
F8-C3 & 04:21:17.10 & +37:35:39.73 & & 0.26 & 0.16 & 75 & -0.41 & 0.39 & 0.53 
& 2.74 \\
F8-C4 & 04:21:37.71 & +37:34:11.82 & & 0.38 & 0.34 & 153 & -1.24 & 0.68 & 2.98 
& 1.63 & Class II\\
\enddata
\tablenotemark{}\\
\tablenotemark{$\ast$}{~Virial parameter, $\alpha_{\rm vir}$, is given by the ratio of $M_{\rm vir} / M_{\rm obs}$.}\\
\tablenotemark{$\dagger$}{~F4-C1, F4-C3, and F8-C4 are identified by \citet{broekhoven-fiene2018} based on the $Spitzer$ and $Herschel$ and its classification given here is based on its IR spectral slope. F4-C3 and F8-C4 are detected in both of $Spitzer$ and $Herschel$, but F4-C1 is only detected in $Herschel$ PACS.} \\
\end{deluxetable*}
\vspace{5mm}

\subsection{Physical Properties of the identified filaments} \label{sec:phypro}

For the ten filaments identified with the dendrogram technique, we derived physical quantities such as molecular gas mass, length and width, mass per unit length, peak velocity distribution, and nonthermal velocity dispersion. \\

\subsubsection{H$_{\rm 2}$ Column Density and Mass} \label{sec:h2mass}

We estimated H$_{\rm 2}$ column density from the C$^{\rm 18}$O column density using the formula \citep{garden1991,pattle2015} :

\begin{equation}
	N = \frac{3 k_{\rm B}}{8 \pi^{\rm 3} B \mu^{\rm 2}} \frac{e^{h B J(J+1)/k_{\rm B} T_{\rm ex}}}{J+1} \frac{T_{\rm ex} + \frac{h B}{3 k_{\rm B}}}{1 - e^{-h \nu / k_{\rm B} T_{\rm ex}}} \int \tau \rm d \it v , \label{eq:eqn}
\end{equation}

\noindent where $B$ is the rotational constant, $\mu$ is the permanent dipole moment of the molecule, $J$ is the lower rotational level, and $T_{\rm ex}$ is the excitation temperature. The excitation temperature is calculated following \citet{pineda2008} : 

\begin{equation}
	T_{\rm ex} = \frac{T_{\rm 0}} {\rm ln ( 1+ \it T_{\rm  0} (\frac{T_{\rm R}}{1-e^{-\tau}} + \frac{T_{\rm 0}}{e^{T_{\rm 0}/T_{\rm bg}} - 1} )^{\rm -1} )}
\end{equation}

\noindent where $T_{\rm 0}=h \nu / k_{\rm B}$ and $T_{\rm bg}$ is the cosmic microwave background temperature (2.73~K). We used the radiation temperature of $\tco$ for $T_{\rm R}$ with assumptions that $\ceo$ and $\tco$ trace materials with the same excitation temperature and $\tco$ is optically thick. We derived optical depths $\tau$ of $\tco$ and $\ceo$ with the abundance ratio of [$\tco / \ceo$] = 5.5 \citep{frerking1982} and the relation of 
\begin{equation}
	\frac{T_{\rm max,\ceo}}{T_{\rm max,\tco}} = \frac{1 - e^{-\tau_{\ceo}}}{1 - e^{-\tau_{\tco}}} ,
\end{equation} 
where $T_{\rm max,\ceo}$ and $T_{\rm max,\tco}$ are the maximum intensities of $\ceo$ and $\tco$, respectively. In regions far from the main filaments, $\tco$ is not necessarily optically thick. In this case we adopted the excitation temperature for their nearest position where $\tau_{\tco} > 1$. The resulted excitation temperature ranges between 5 and 11~K. \citet{imara2017} derived temperature from $^{12}$CO data and provided $\sim$ 7~K for F6 and F7. Our $\tex$ of these two filaments ranges about 5 to 9.5~K and the average $\tex$ is $\sim$7~K which is well matched with that of \citet{imara2017}.

The rightmost integration term of Equation~1
 can be written as \citep{pattle2015} :
\begin{eqnarray*}
	\int \tau (v) \rm ~d \it v &=& \frac{1} {J(T_{\rm ex})-J(T_{\rm bg}) } \int \frac{\tau (v)}{1- e^{- \tau (v)}} ~T_{\rm mb} \rm ~d \it v \\
	&\approx & \frac{1}{J(T_{\rm ex})-J(T_{\rm bg})} \frac{\tau (v_{\rm 0})}{1 - e^{- \tau (v_{\rm 0})}} \int T_{\rm mb} \rm ~d \it v ,	
\end{eqnarray*} 
\noindent where $v_{\rm 0}$ is the central velocity, $T_{\rm mb}$ is the observed main beam temperature of the line and $J(T)$ is the source function, $J(T) = T_{\rm 0} / ( \rm e^{\it T_{\rm 0} / \it T} -1)$ and $T_{\rm 0} = h \nu / k_{\rm B}$, and $T_{\rm ex}$ and $T_{\rm bg}$ are the excitation temperature and the cosmic microwave background temperatures described as in the Equation~1
. For calculation of $ \int T_{\rm mb} \rm d \it v$, we used the area under the fitted gaussian function. Since most of the spectra have a shape of the Gaussian profile, the difference between the area under the fitted Gaussian function and the integrated $T_{\rm mb}$ within the velocity range of $v_{\rm cen} \pm \triangle v/2 $, where $v_{\rm cen}$ and $\triangle v$ are the central velocity and linewidth of Gaussian fitting results, respectively, is found to be less than 5\%.

$\nht$ from $N_{\rm \ceo}$ is calculated with the conversion factor of $\nht / N_{\rm ^{12}CO} = 1.1 \times 10^{4}$ \citep{pineda2010} and the abundance ratios of $\rm ^{12}CO / \tco = 69$ \citep{wilson1999} and $\tco / \ceo = 5.5$ \citep{frerking1982}. The derived H$_{2}$ column density distribution is shown in the panel (b) of Figure~6
.
\begin{figure*}
\includegraphics{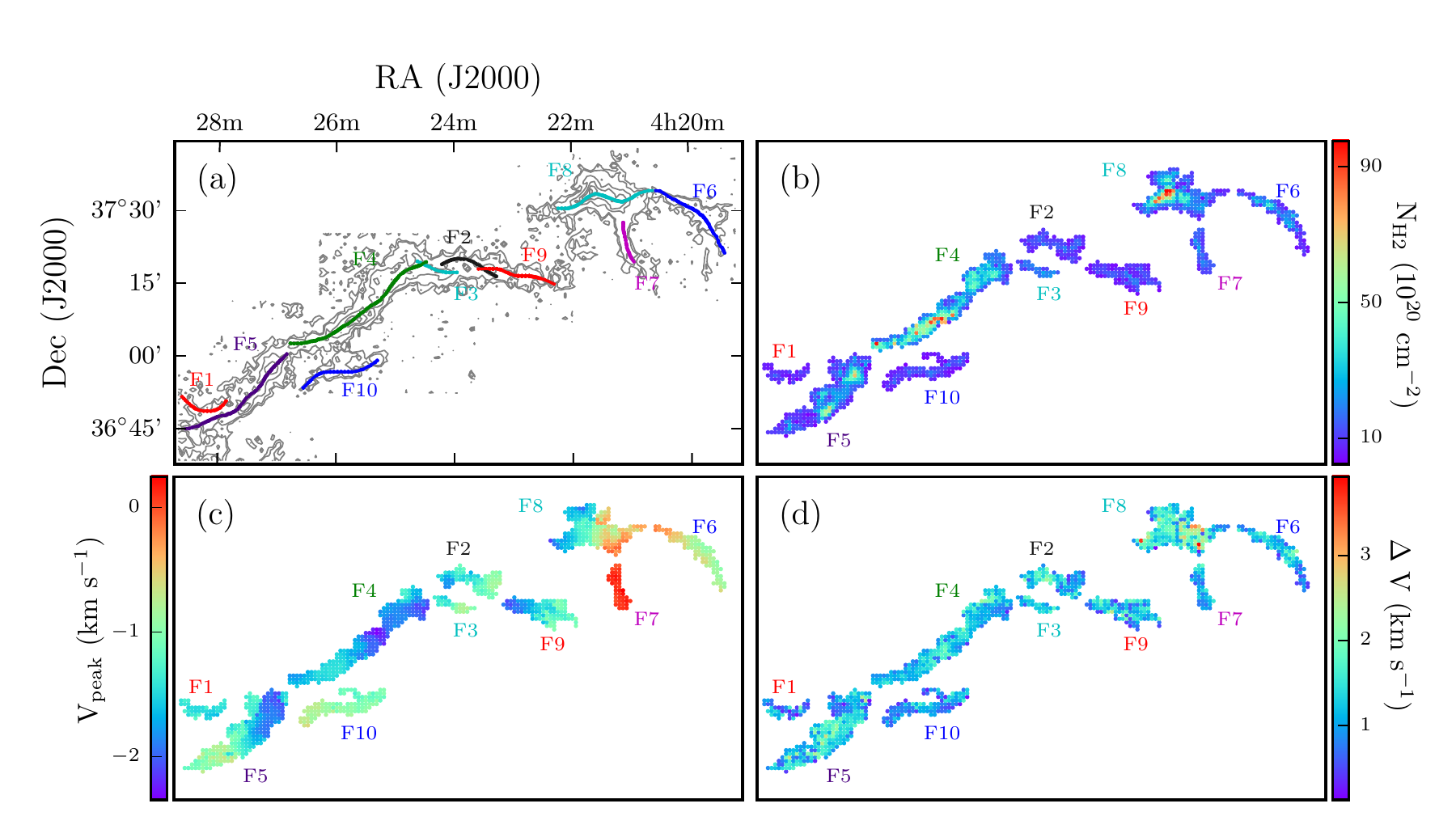}
\caption{(a) Locations of ten identified filaments (skeletons) on top of the integrated intensity map of $\ceo$ (the same as the contour map in the bottom panel of Figure~2
; contours are 2, 3, 6, 9, $\cdots, 18 \times \sigma$ in K~$\kms$). (b) - (d) H$_{2}$ column density, velocity field, and linewidths maps of each filament. $V_{\rm peak}$ and $\Delta V$ (linewidth) are derived quantities by Gaussian fitting method. Small offset is given to the original position of each filament to avoid spatial overlaps and distinguish from each other.
\label{fig:ivdvmaps}}
\end{figure*}

We also estimated the H$_{\rm 2}$ column density from $Herschel$ data to check whether the value derived with $\ceo$ is reliable or not. Firstly, 250~$\mu$m, 350~$\mu$m, and 500~$\mu$m $Herschel$ data were convolved to the 44$^{\prime \prime}$ pixel size and then co-aligned on the TRAO $\ceo$ and $\tco$ pixel-grid. Secondly, we derived spectral energy distribution (SED) fits from 250~$\mu$m, 350~$\mu$m, and 500~$\mu$m data for each pixel position with simple dust emission model, $F_{\nu} = \kappa_{\nu} \times B(\nu, T) \times$~column density. A dust opacity law of $\kappa_{\nu} = 0.1(\nu /1000 \rm GHz)^{\beta} cm^{2} g^{-1}$ is assumed with fixed dust emissivity index $\beta$ of 2 \citep{draine1984,schnee2010} and the standard mean molecular weight, $\mu$, of 2.86 is used for the calculation of H$_{2}$ column density \citep{kauffmann2008} . 
\begin{figure}
\plotone{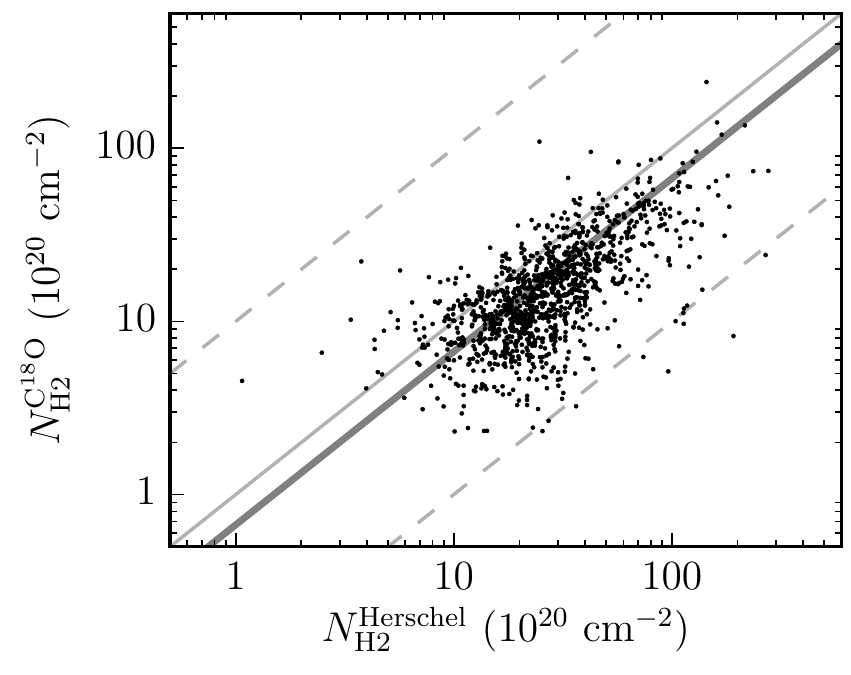}
\caption{Comparison of H$_{\rm 2}$ column densities from $\ceo$ and $Herschel$ data. The solid thin gray line superimposed indicates where $\nht ^{\ceo}$ and $\nht ^{Herschel}$ are identical, and the two dashed lines show where the ratio of $\nht ^{\ceo} / \nht ^{Herschel}$ is 10 and 0.1. The thick solid line presents the least squares fit result. \label{fig:nh2comp}}
\end{figure}

The measured H$_{\rm 2}$ column densities from $\ceo$ and $Herschel$ data are compared in Figure~7
. It appears that $\nht^{\ceo}$ is smaller than $\nht^{Herschel}$ by a factor of $\sim$2, but with good linear correlation. 

\begin{deluxetable*}{c r@{~to~}l c c c c c c c r@{$\pm$}l ccc}
\tablecaption{Physical Properties of Filaments \label{tab:ppfila}}
\tablecolumns{15}
\tablewidth{0pt}
\tablehead{
\colhead{Fil. ID} &
\multicolumn2c{$V_{\rm peak}$ range} &
\colhead{} &
\colhead{$\bar \sigma_{\rm tot}$} &
\colhead{$L$} &
\colhead{$W$} & 
\multicolumn2c{$M$} & 
\colhead{$\mlin$} & 
\multicolumn2c{ $\mlin^{\rm crit}$} & 
\colhead{$| \nabla V_{\rm peak} |$} &
\colhead{cores} &
\colhead{YSOs} \\
\colhead{(1)} & \multicolumn2c{(2)} & \colhead{} & \colhead{(3)} & \colhead{(4)} & \colhead{(5)} & \multicolumn2c{(6)} & \colhead{(7)} & \multicolumn2c{(8)} & \colhead{(9)} & \colhead{(10)} & \colhead{(11)} 
}
\startdata
 1 & -1.7 & -1.2 &  & 0.18 & 0.45 &  0.10  &   6.9 &  &  15.3 &  14.8 & 12.3 & 1.12 & N/A &  \\
 2 & -1.9 & -0.8 &  & 0.24 & 0.51 &  0.12  &  17.8 &  &  34.7 &  25.6 & 17.4 & 1.90 & N/A & \\
 3 & -1.6 & -0.8 &  & 0.24 & 0.35 &  0.07  &   9.4 &  &  26.5 &  27.6 & 12.1 & 2.61 & N/A &  \\
 4 & -2.3 & -1.3 &  & 0.24 & 1.40 &  0.08  & 216.1 &  & 154.1 &  25.7 & 12.9 & 0.69 & 3 & 2 \\
 5 & -2.2 & -0.6 &  & 0.25 & 1.12 &  0.19  &  85.6 &  &  76.5 &  28.3 & 19.7 & 1.28 & 1 & \\
 6 & -1.0 & -0.2 &  & 0.22 & 0.83 &  0.10  &  25.7 &  &  30.9 &  22.0 & 14.7 & 0.89 & N/D & \\
 7 &  0.0 &  0.3 &  & 0.21 & 0.35 &  0.13  &  12.3 &  &  34.9 &  20.0 & 10.3 & 0.63 & N/D & \\
 8 & -2.1 & -0.1 &  & 0.31 & 0.87 & $\cdots$ & 127.8 &  & 147.6 &  45.0  & 33.7 & 2.26 & 4 & 1 \\
 9 & -2.1 & -1.0 &  & 0.23 & 0.66 &  0.14  &  20.0 &  &  30.1 &  23.5 & 17.5 & 1.57 & N/A & \\
10 & -1.3 & -0.6 &  & 0.18 & 0.70 &  0.08  &  19.3 &  &  27.5 &  14.9 &  9.7 & 0.85 & N/A & \\
\enddata 
\tablecomments{Col.(1) Filament ID. Col.(2) The largest and smallest $\vpeak$ in $\kms$. Col.(3) Averaged total velocity dispersion from the $\ceo$ linewidths in $\kms$. Col.(4) Length of filament measured from the eastmost (or northmost) point to the westmost (or southmost) point of skeleton in pc. Col.(5) Filament width in pc, i.e., FWHM of radial profile of H$_{2}$ column density. Col.(6) H$_{2}$ mass of filament in $M_{\odot}$. Col.(7) Mass per unit length of filament in $M_{\odot} ~ \rm pc^{-1}$. Col.(8) Effective critical mass per unit length of filament derived with the mean total velocity dispersion in $M_{\odot}~ \rm pc^{-1}$ (see Section~\ref{sec:grav}). Col.(9) Mean velocity gradient of filament in $\kms ~ \rm pc^{-1}$. Col.(10) Number of dense cores identified with $\nthp$ data. F1, F2, F3, F9, and F10 are not observed with $\nthp$. F6 and F7 are observed with $\nthp$ but no emission is detected at the $rms$ level of 0.06~K$\rm [T_{\rm A}^{\ast}]$. Col.(11) YSOs identified with $Spitzer$ and $Herschel$ \citep{broekhoven-fiene2018}. }
\end{deluxetable*}

The multiple velocity components of filaments can cause $\nht^{\rm \ceo}$ to be smaller than $\nht^{\rm Herschel}$. The spectra with multiple velocity components are presented in Figures~2
 and the Appendix (Figure~14 and 15
 ), and multiple velocity components of filaments are overlapped in the plane of sky. In this area, H$_{\rm 2}$ column density from $\ceo$ is derived separately for each filament components, while $\nht$ from $Herschel$ dust emission is summed for the different filaments. Hence $\nht^{\ceo}$ becomes less than $\nht^{\rm Herschel}$. We should also add the other well known facts the CO depletion and dissociation that can cause differences between H$_{2}$ column densities derived with the two methods. At high densities of $n_{\rm H_{2}} \gtrsim 10^{5}~\rm cm^{-3}$ and low dust temperatures of $< 20$K, CO is depleted through freeze-out onto dust grains. Besides, Herschel is sensitive to the large scale structure of the cloud, where the dust is warmer and CO may be, at least partially, photodissociated, so not properly tracing this outer zone of the cloud \citep[e.g.,][]{caselli1999,tafalla2004,difrancesco2007,spezzano2016}.

The difference between $\nht^{\rm \ceo}$ and $\nht^{\rm Herschel}$ can be also affected by the use of some uncertain parameters. The typical error of H$_{2}$ column density measured from $Herschel$ data is a factor of 2, which is mainly caused by the uncertainty of the dust opacity law. In case of $\nht$ from $\ceo$, it is much complicated because the conversion factor of CO-to-H$_{2}$ and the ratios of $\rm CO / \tco$ and $\tco / \ceo$ vary by a factor of up to 5 according to the metallicity, column density, and temperature gradients \citep[see][and references therein]{pineda2010,bol13}. Hence, the uncertainty of $\nht$ measured from $\ceo$ is likely to be a factor of a few, and we can conclude that the resulted $\nht^{\rm \ceo}$ and $\nht^{\rm Herschel}$ is well matched with each other within a range of uncertainty.

H$_{\rm 2}$ column density measured with $\ceo$ ranges $2 - 100 \times 10^{20} \rm ~cm^{-2}$, and the H$_{\rm 2}$ mass of each identified filament is $\sim$10 to 200~$M_{\odot}$. The estimated mass of each filament is tabulated in Table~3
.  \\

\subsubsection{Length and Width} \label{sec:lengths}

We estimated the filament's length along the skeleton without any correction of the projection effect. Filament skeletons are determined in the following procedures. First, we made moment 0 images for each filament, then applied \textsc{FilFinder} which uses Medial Axis Transform method to the moment 0 images for each filament. Medial Axis Transform method gives a skeleton which is the set of central pixels of inscribed circles having maximum radius \citep{koch2015}. The skeletons determined with this process are well matched with the $skeleton$ obtained with \textsc{DisPerSE} (DIScrete PERsistent Structures Extractor). The \textsc{DisPerSE} algorithm finds the skeleton by connecting the critical points of maxima and/or saddle points which has a zero gradient in the map and identifies skeletons of filaments \citep{sousbie11}. 
There is a slight deviation among the skeletons obtained between two algorithms which is less than 2 pixel size in most cases. The skeletons of identified filaments with their IDs are shown in the panel (a) of Figure~6
. Lengths of filaments in L1478 range from $\sim 3^{\prime}$ to 11$^{\prime}$ which correspond to about 0.4~pc to 1.4~pc at the distance of 450~pc. 

The filament width is measured from the H$_{2}$ column density map (see section~3.3.1
 and Figure~6
) of each filament. We made radial profile of H$_{2}$ column density versus distance from the skeleton, applied gaussian fitting, and derived the FWHM as filament's width.

The resulted lengths and widths of filaments are listed in Table~3
. As can be seen, the width ranges from $\sim$0.1 to 0.2 pc. This is consistent with the typical filament width of 0.1~pc which is shown in $Herschel$ images~\citep{arzoumanian2011}. But it should be considered here that our spatial resolution of $\ceo$ is $\sim 47^{\prime \prime}$, corresponding 0.1~pc at the distance of 450~pc and thus our result on the width may be affected by our observing resolution in space. \\

\subsubsection{Mass per unit length}

With the mass and length of filament, we calculated the mean of the mass per unit length, $\mlin$. We simply divided the mass of filaments with its lengths given in Section~3.3.1 and 3.3.2
 and thus $\mlin$ shown here is the averaged mass per unit length of the filament. The mass per unit length of the filaments in L1478 is estimated to range from $\sim$20 to 150 $M_{\odot}~ \rm pc^{-1}$. \\ 

\subsubsection{Global Velocity Field}

Intensity weighted mean velocity (moment 1) and velocity dispersion (moment 2) maps are very useful for understanding of the global velocity properties of molecular clouds, but may not be appropriate to study the kinematics of filaments showing multiple velocity components due to their spatial overlaps. Hence, we carried out Gaussian fitting for $\ceo$ spectra to extract velocity information of the filaments. 

\begin{figure*} \plotone{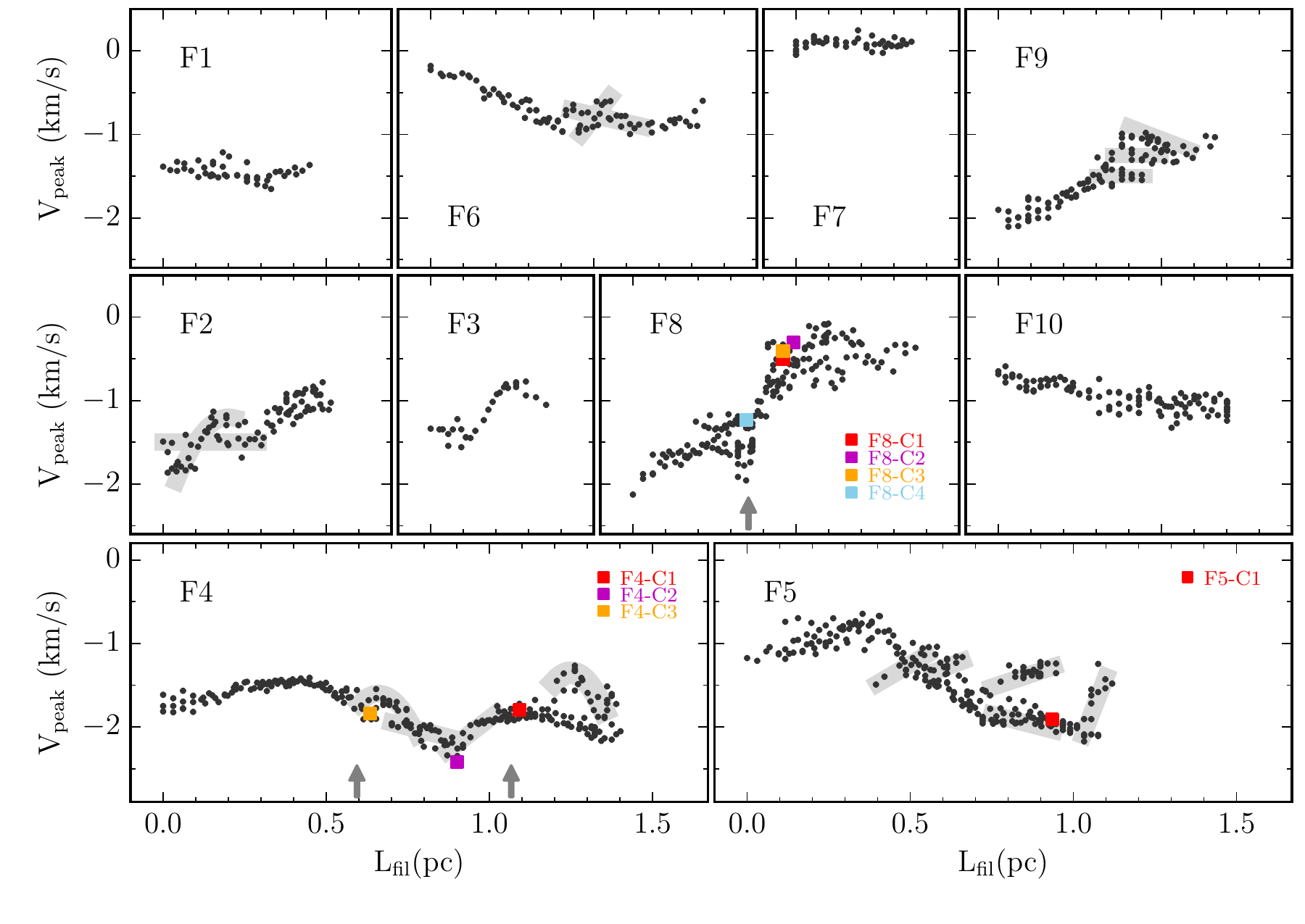} \caption{Velocity structures along the filaments. $L_{\rm fil}$ is calculated along the skeleton points (\S~3.1) 
from the eastmost point of each filaments. $\vpeak$ from $\ceo$ is presented with solid black dots. $\vpeak$ of each core denoted here is derived from the averaged $\nthp$ (see Figure~4 and 5). 
The gray thick lines are drawn to highlight some different velocity components. The YSOs reported by \citet{broekhoven-fiene2018} are presented with gray arrows in their positions along the skeleton. \label{fig:lfil_vpeak}} \end{figure*}

The global continuous velocity distribution and gradient in the filaments can be seen in the peak velocity distribution map, panel (c) in Figure~6
. In section~3.1
, we mentioned that F1 to F5 are from one stem though they are found as an independent leaf, and this is also true for F6 to F8. Those filaments are identified as an independent filament because they have a maximum of 2$\sigma$ higher than the saddle point and they are all connected to each other. The peak velocity map shows that they are spatially and kinematically continuous. Especially, it is shown that F6 and F7 are continuously connected with F8 in the velocity space as well as in the plane of sky, indicating they are a clear hub-filaments structure, i.e., dense star-forming hubs with multiple filaments. 

Information of the velocity distribution of each filament is tabulated in Table~3
 as quantities of $\vpeak$ ranges and gradients ($| \bigtriangledown \vpeak |$) of filaments. Figure~8
  shows the peak velocity distribution along the skeleton of each filament. The slope (velocity gradient) is different from filament to filament, and seems to be intrinsic to each filament, although no inclination correction has been applied. The variation of $\vpeak$ ($\triangle \vpeak$) changes between $\sim 1~ \kms$ and 3 $\kms$ for each filament. F5 and F8 have the largest $\triangle \vpeak$ of $\sim 2~\kms$ and F7 has the smallest $\triangle \vpeak$ of 0.3~$\kms$. The mean velocity gradient ($| \bigtriangledown \vpeak |$) ranges from about 0.6 to 2.6~$\kms ~pc^{-1}$ and F3 shows the largest $| \bigtriangledown \vpeak |$ of $\sim 2.6~ \kms~ \rm pc^{-1}$.

We can see two clear kinematic properties of filaments. The one is the coherence of velocities of filaments, i.e., every filament, even F8 which has a hub-filaments shape, has continuous $\vpeak$ along the skeletons. The other one we notice is that some different velocity components appear (drawn with thick gray lines for F2, F4, F5, F6 and F9). We will discuss more about the different velocity components in Section~4. \\ 

\subsubsection{Nonthermal velocity dispersion} 

Nonthermal velocity dispersion ($\sigma_{\rm NT}$) can be calculated as following: \begin{equation} \sigma_{\rm NT} = (\sigma^{2} - \frac{k_{\rm B} T}{\mu m_{\rm H}} )^{1/2} \end{equation} where $\sigma$ is velocity dispersion derived from the FWHM, $k_{\rm B}$ is the Boltzmann constant, $T$ is the gas temperature, $\mu$ is the mean molecular weight of $\ceo$ and $\nthp$, and $m_{\rm H}$ is the mass of the atomic hydrogen. 

The FWHM for $\ceo$ and $\nthp$ were obtained from Gaussian fit to those spectra. In case of $\nthp$ spectra seven hyperfine components were simultaneously fitted with seven Gaussian forms at once using their line parameters given by \citet{caselli1995}. The gas temperature is assumed to be the same as $T_{\rm ex}$ which is derived in Section~3.3.1.

\begin{figure*} \plotone{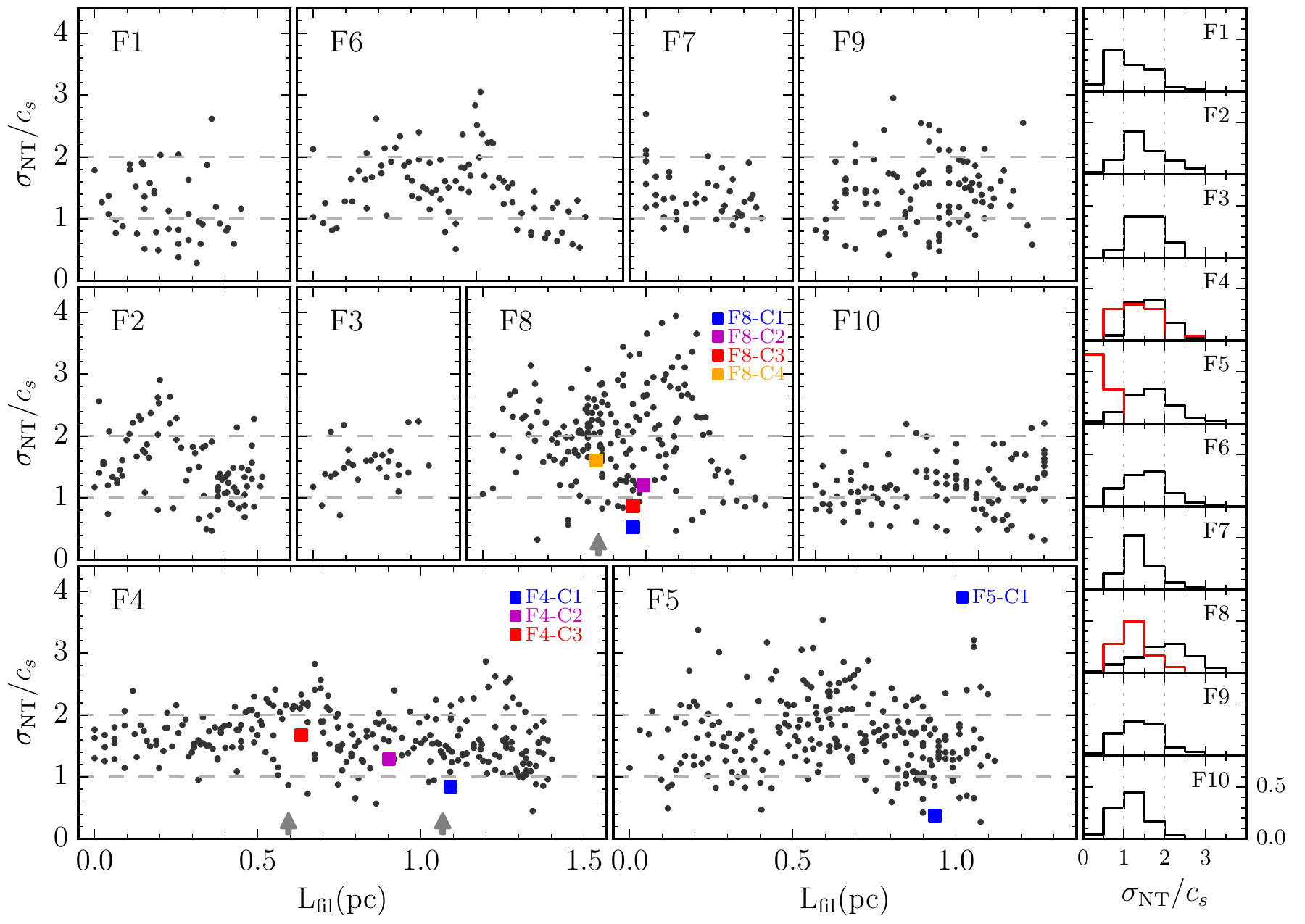} \caption{Velocity dispersions in all identified filaments and dense cores. {\it Left panels:} Nonthermal velocity dispersions normalized by the local sound speed as a function of the position along each filament. The $\sigma_{\rm NT}/ c_{\rm s}$ derived from $\ceo$ is denoted with solid black dots. $\sigma_{\rm NT}/ c_{\rm s}$ derived from the averaged $\nthp$ spectrum of each dense core (see Figures~4 and 5
) are presented with different colors of squares. The positions of YSOs reported by \citet{broekhoven-fiene2018} are presented with gray arrows. {\it Right panels:} Normalized histograms of $\sigma_{\rm NT} / c_{\rm s}$. The $\sigma_{\rm NT} / c_{\rm s}$ derived from linewidths of $\ceo$ and of $\nthp$ are presented with blue and red, respectively. In here, histograms of $\sigma_{\rm NT}/ c_{\rm s}$ derived from $\nthp$ are not from the averaged spectrum but from every detected $\nthp$ spectra (S/N $> 5$). \label{fig:lfil_ntvcs}} \end{figure*}

Figure~9 
shows the nonthermal velocity dispersions in our all identified filaments and dense cores. Except F8, $\sigma_{\rm NT} / c_{\rm s}$ of every filament peaks at transonic regime ($\sim$1.5). Though F1 has a peak of $\sigma_{\rm NT} / c_{\rm s}$ between 0.5 and 1.0, the portion of transonic nonthermal velocity dispersion is about the same as that of subsonic components (47\% each). F8 that has YSOs appears to have $\sim$51\% of spectra of supersonic velocity dispersions. 

Nonthermal velocity dispersions measured from the dense core tracer $\nthp$ (denoted with squares of various colors) are smaller than those from $\ceo$, either subsonic or transonic. The histogram of $\sigma_{\rm NT} / c_{\rm s}$ from $\nthp$ (presented with red color) shows that $\sigma_{\rm NT} / c_{\rm s}$ from $\nthp$ of F4 and F8 peaks at transonic regime ($1 \lesssim \sigma_{\rm NT}/ c_{\rm s} \lesssim 2 $), while that of F5 peaks at subsonic values. F4 shows similar velocity dispersions in filaments traced by $\ceo$ and in dense cores traced by $\nthp$, but F5 and F8 have different distributions of $\sigma_{\rm NT}^{\ceo}$ and $\sigma_{\rm NT}^{\nthp}$. \\

\subsubsection{Chemical Evolution of Dense Cores}

Integrated intensity maps of $\nthp$, $\ceo$, and SO molecular lines are presented in Figure~10 
with the dense cores identified in \S~3.2. 
SO is known as one of the most sensitive molecules of molecular depletion and $\nthp$ is known to survive much longer at high density when CO becomes depleted in gas phase \citep[e.g.,][]{tafalla2006}. 

\begin{figure*}[ht!]
\plotone{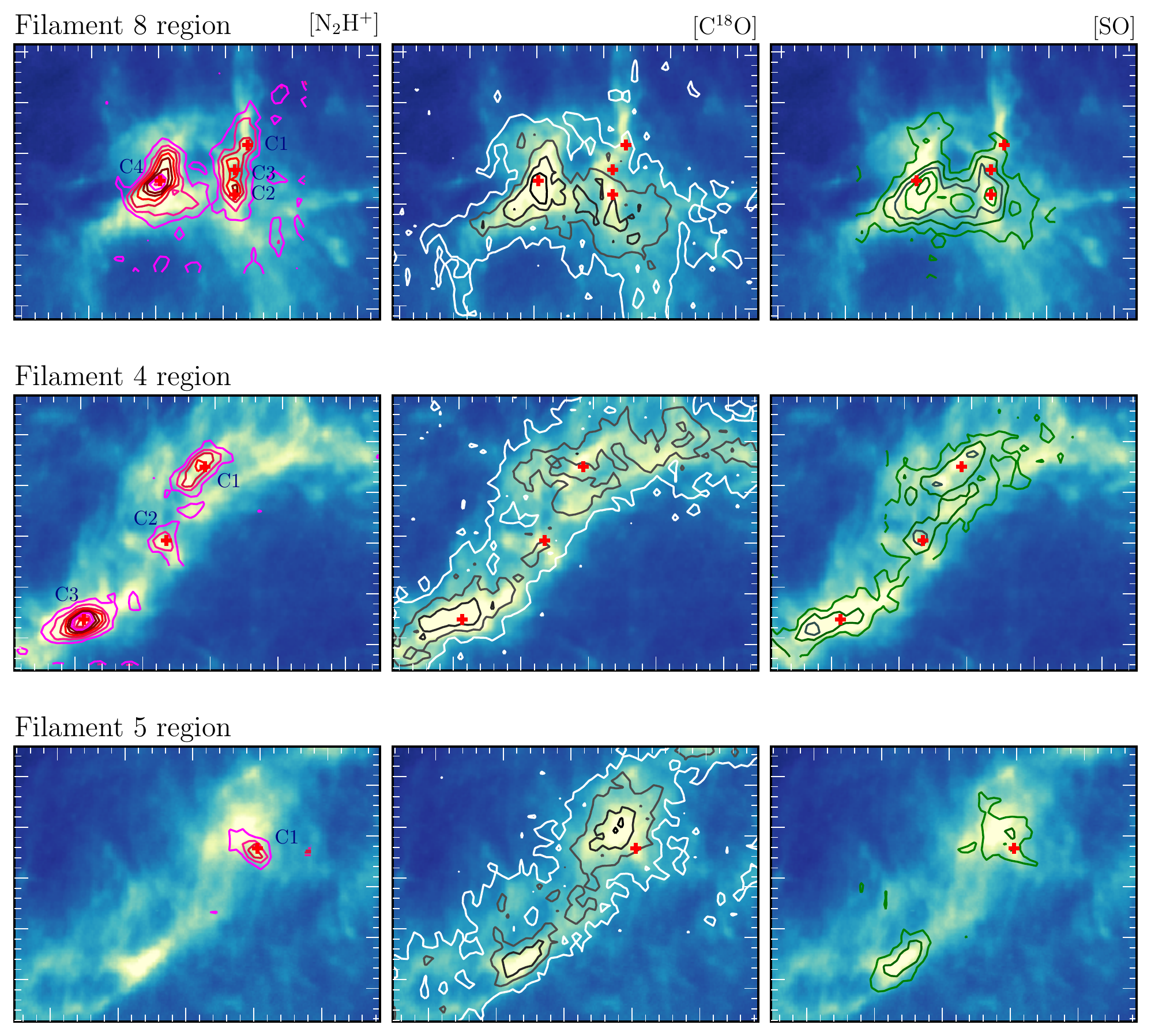}
\caption{Integrated intensity contours maps of $\nthp$ (left), $\ceo$ (middle), and SO (right) on the Herschel 250$\mu$m background image. Contours are at every 3$\sigma$ interval from 3$\sigma$. The $\nthp$ peak positions of dense cores identified are presented with red crosses. \label{fig:n2hpcsso}}
\end{figure*}

As can be seen in the maps, $\nthp$ traces more compact regions than $\ceo$ and SO. Chain-like structure of cores 
can be seen in F8 and F4 regions. F8-C1, F8-C2, and F8-C3 and F4-C1, F4-C2, and F4-C3 stand in lines and their $\nthp$ emission contours show chain-like structure which is, however, not clear in $\ceo$ and SO because of their possible depletion. 

In the F8 region (top panels), four dense cores are identified. F8-C1 and -C3 show relatively weak emissions in the three molecular lines of $\nthp$, $\ceo$, and SO. Meanwhile, F8-C2 has a SO condensation at the peak position of $\nthp$. F8-C4, which is the largest and brightest core, contains a class II object. Hence, F8-C4 can be the most evolved core in this filament and F8-C2 seems to be relatively younger than F8-C4. 

In the F4 region (middle panel), three dense cores are found. $\ceo$ is extended from the southeast to northwest, but there is no $\nthp$ and SO emission toward northwest. F4-C3 is the largest core with the brightest and the centrally peaked $\nthp$ emission in the F4 region, but $\ceo$ and SO is not as peaked as $\nthp$. This is the same for F4-C1 which has centrally peaked $\nthp$ but not $\ceo$ and SO. On the contrary, F4-C2, which has the weakest $\nthp$ emission, has a similarly peaked in three lines. Hence, F4-C2 appears to be the youngest core in the Filament 4 region. 

In F5 region (bottom panels), only one core is found. Peaks of $\ceo$ and SO show slight offset from the $\nthp$ peak, and thus can be depleted. This indicates that F5-C1 is not a young but evolved core. On the other hand, at the southeastern region of F5-C1, where no $\nthp$ emission is detected, $\ceo$ and SO emissions are clearly detected indicating that this region is at an earlier stage. 

The chemical differences between cores which form in a filament indicate that they are at different stage of evolution. \citet{lee2011} carried out various molecular line observations toward several tens of starless cores, and found that the column density increases in a sequence of core evolution that the earliest stage is static cores, and the next is expanding and/or oscillating cores, and the most evolved one is contracting cores. To probe the relation of column density and the evolution status, we compared the H$_{2}$ column density between the cores. The youngest cores in F4 and F8 are F4-C2 and F8-C2, respectively, and they have lower H$_{2}$ column densities than other dense cores, which agrees the result of \citet{lee2011} that cores with higher H$_{2}$ column density tend to be more evolved than others. However, the internal motions such as static, expanding and/or oscillating, and contracting motions in each evolutionary stage are not clearly related with the evolutionary stage in our dense cores. In Filament 4, the youngest core F4-C2 doesn't show clear infall signature, but the other cores of F4-C1 and -C3 show clear infall motions indicating the contraction of cores. On the contrary, in Filament 8, the most evolved core F8-C4 doesn't show any infall signature, but F8-C1 and -C2 which are younger than F8-C4 show blue asymmetry showing contraction. However, because of our small number of dense cores, it is difficult to conclude that the internal motions of cores are not correlated with the evolutionary stage of cores. \\

\section{Discussion} \label{sec:disc}

\subsection{Fibers, building blocks of filaments?} 

Recent molecular line observations of filaments with higher spatial and spectral resolutions lead us to confirm the presence of fibers. \citet{hacar2018} proposed that all filaments are bundles of fibers which have a typical length and width of $\sim$0.5~pc and 0.035 to 0.1~pc, respectively, and are velocity-coherent and characterized by transonic internal motions. Fibers are found in various star-forming environments, from low to high-mass star-forming clouds \citep[e.g.,][]{hacar2013,hacar2016,hacar2017,maureira2017,clarke2018,dhabal2018,hacar2018}. Interestingly, \citet{hacar2018} showed that the mass per unit length of the total filament has a linear correlation with the surface density of fibers and suggested an unified star-formation scenario for the isolated, low-mass and the clustered, high-mass stellar populations where the initial number density of fibers may determine the star forming properties of molecular clouds and filaments. Our beam size ($\sim$0.1~pc at the distance of 450~pc) is comparable to the width of the fibers and it is smaller than the typical lengths of fibers. Besides, our velocity resolution is enough to check the presence of fibers.

In Figure~8, 
the $\vpeak$ distribution along the skeleton appears to be coherent in each filament, but fiber-like velocity structures can be seen in every filament, especially in F2, F4, F5, and F6. F4, which has YSOs and dense cores, seems to have fibers weaving together at positions of $\lfil \sim$0.6, 0.9, and 1.2~pc (indicated with gray thick curves in the Figure~9). 
It is interesting that the junction of fibers with slightly different velocity is consistent with the position of dense cores. F5 looks like having one long main filament and shorter fibers are interlaced with the main filament. Actually, at the northwestern part of F5 (at $\lfil \sim$ 0.7 to 1.0), it shows a clear double peak in $\ceo$ spectra and the difference of peak velocities is $\sim 0.5~ \kms$. In F2, F6, and F9, the braided fibers can be found, too. These three filaments seem to be composed of two parallel fibers which meet each other at the center of the filament. F8 has hub-like morphology, but its velocity structure also shows the presence of fibers.

What we can see with our data is still limited because of its poor spatial resolution, but one noticeable thing is that the filaments having YSOs and dense cores show clear evidence for presence of fibers than other filaments do. F4 and F5 look like to have more fibers than other filaments in L1478, and furthermore they are gravitationally unstable and have YSOs and dense cores. In other words, the peak velocity distributions of filaments in L1478 indirectly indicate that filaments are made of fibers and those filaments gravitationally collapsing show denser and finer distribution of fibers. These results support the star formation scenario suggested by \citet{hacar2018} that filaments are bundles of velocity-coherent fibers and fibers are an arbiter of star formation in molecular clouds in a sense. \\

\subsection{Are the filaments in L1478 gravitationally bound?} \label{sec:grav}

The ubiquity of filaments indicates that the filament is an indispensable structure in the star formation process. \citet{andre2010} proposed a core formation scenario based on the $Herschel$ Gould Belt Survey observation that long and thin filaments are made first in the molecular clouds and then prestellar cores form by hierarchical fragmentation of gravitationally unstable filaments. The equilibrium line mass or mass per unit length for an isothermal, unmagnitized filament in pressure equilibrium is $M_{\rm line,eq}^{\rm unmag} = 2 c_{\rm s}^{2} / G$ where $c_{\rm s}$ is the isothermal sound speed \citep{ostriker1964,inutsuka1992,inutsuka1997}. The critical value at 10~K is $\sim 15~ M_{\odot}~\rm pc^{-1}$, and a filament with larger $\mlin$ than $M_{\rm line,eq}$ becomes supercritical and undergoes gravitational contraction. 

Filaments in L1478 have mass per unit length of $\sim$20 - 150 $M_{\odot}~ \rm pc^{-1}$ that is larger than the equilibrium mass per unit lengths. This means that every filaments in L1478 is supercritical and will collapse gravitationally, and dense cores can be formed by gravitational fragmentation along the filaments. However, in this case, nonthermal components such as turbulence motions as well as magnetic field are not considered. Hence, we derived the effective critical mass per unit length which includes nonthermal motions, $\mlin ^{\rm crit} = 2 {\bar \sigma_{\rm tot}}^{2} / G$, where $\bar \sigma_{\rm tot}$ is the average total velocity dispersion of a filament \citep[e.g.,][]{arzoumanian2013,peretto2014}. The calculated $\mlin ^{\rm crit}$ with ${\bar \sigma_{\rm tot}}$ is tabulated in Table~3 
and presented in panel (a) of Figure~11 
with thick gray bar. The ratio, $\mlin / \mlin ^{\rm crit}$, is given in panel (b) of Figure~11. 

\begin{figure}
\plotone{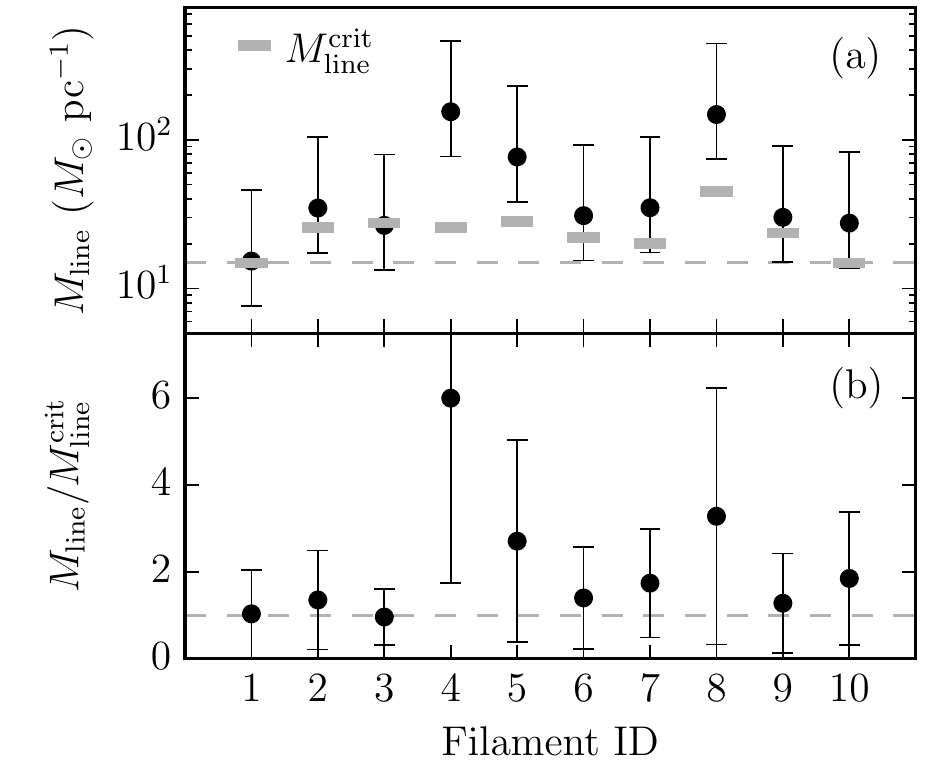}
\caption{Criticality of observed filaments. (a) The mass per unit length ($M_{\rm line}$) is presented with solid dots and the effective critical mass per unit length derived with the average $\sigma_{\rm tot}$ for each filament is presented with a gray bar. The gray dashed line denotes the equilibrium value ($\sim 15~M_{\odot} ~\rm pc^{-1}$) for isothermal cylinder in pressure equilibrium at 10~K. The error bars indicate the factor of two uncertainties of $\mlin$. (b) Ratios of line mass to effective critical mass per unit length. The gray dashed line indicates the line where $\mlin$ and $\mlin^{\rm crit}$ are identical. \label{fig:filmlin}}
\end{figure}

With $\mlin ^{\rm crit}$, all filaments in L1478 are close to critical and F1, F3, and F9 are marginally critical. The three filaments, F4, F5, and F8, are highly critical. This is consistent with the fact that the high density tracer of $\nthp$ is detected and YSOs are found in these filaments only \citep{broekhoven-fiene2018}. 

To see whether there are any inward motions from the gravitationally bound filaments to the dense core, we check the CS (2-1) and $\nthp$ (1-0) line profiles. Blue asymmetry in spectral profiles is generally used to identify infall signatures of dense cores and molecular clouds \citep[e.g.,][]{leung1977,lee2001}. In centrally concentrated cores, the foreground gas appears to be absorbed in optically thick lines. If inward motions prevail, the absorption of foreground gas becomes redshifted and the optically thick emission line appears to be brighter in blue peak. As shown in Figure~12, 
among the eight dense cores found from $\nthp$ data, F4-C1, F4-C3, F5-C1, and F8-C2 show the blue asymmetry in the CS (2-1) line profile implying the existence of gas infalling motions which is consistent with their status of criticality. \\

\begin{figure*} \plotone{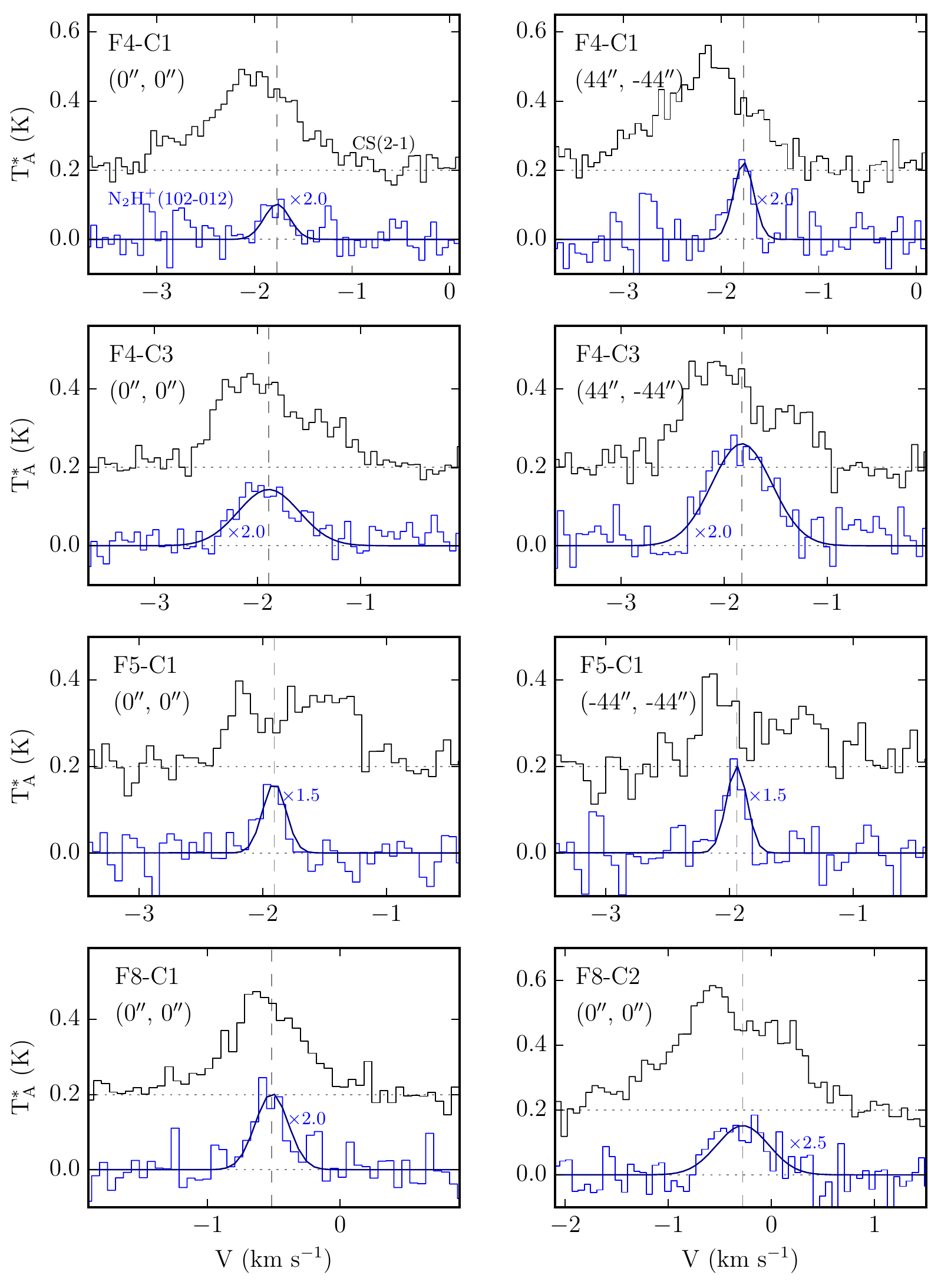} \caption{Infall asymmetric profiles toward dense cores in the supercritical filaments F4, F5, and F8. Averaged CS(2-1) and $\nthp$ (102-012) spectra toward dense cores are drawn with black and blue histograms, respectively. Offsets from the core center are given in the top left corner of each panel. The blue solid curve and gray dashed vertical lines are the hyperfine fitting line and central velocity of cores, respectively. 
\label{fig:infall}} \end{figure*}

\subsection{Do cores form by collisions of turbulent flows?}

Turbulence motions in filaments and dense cores are important to the formation of filaments and dense cores. Numerical simulations of supersonic turbulence gave a result generating dense stuructures like sheets, filaments, and cores \citep[e.g.,][]{padoan2001,xiong2017,haugbolle2018}. \citet{padoan2001} proposed that filaments can form by collision of turbulent flows, and dissipation of turbulence makes the dense cores. Following this colliding model, the filaments and the dense cores may have supersonic and subsonic motions, respectively. To probe this dense core formation scenario, we checked the kinematic properties of filaments and dense cores.

First of all, as shown in Figure~8, 
the $\vpeak$ distribution between dense cores and surrounding material of filaments in F4, F5 and F8 is well matched with each other. This can be seen more clearly in the left panel of Figure~13. 
Peak velocity and velocity dispersion of dense cores and surrounding filament gas were derived with $\nthp$ and $\ceo$ spectrum, respectively. In the left panel of Fig~8, 
the peak velocity of dense core is plotted with that of surrounding filament gas. Every dense core has identical peak velocity to that of filamentary material within the total velocity dispersion traced by $\ceo$. Hence, we can conclude that there is no systemic velocity drift between the dense core and the surrounding material. This result is similar to the previous studies in which no systemic velocity difference is found between the dense cores and surrounding gas \citep[e.g.,][]{kirk2007,hacar2011,punanova2018}.

The nonthermal velocity dispersions of filaments and dense cores are presented in sound speed ($c_{\rm s}$) scale in the right panel of Figure~13. 
As shown in Fig~9, 
$\sigma_{\rm NT}^{\ceo}$ is larger than $\sigma_{\rm NT}^{\nthp}$ for every dense cores. There are four dense cores which have transonic $\sigma_{\rm NT}^{\nthp}$ (F4-C2, F4-C3, F8-C2, and F8-C4), and two dense cores which have subsonic $\sigma_{\rm NT}^{\nthp}$ (F5-C1 and F8-C1). The other two dense cores (F4-C1 and F8-C3) have $\nthp$ velocity dispersions in the borderline between the transonic and subsonic.

\begin{figure*} \plotone{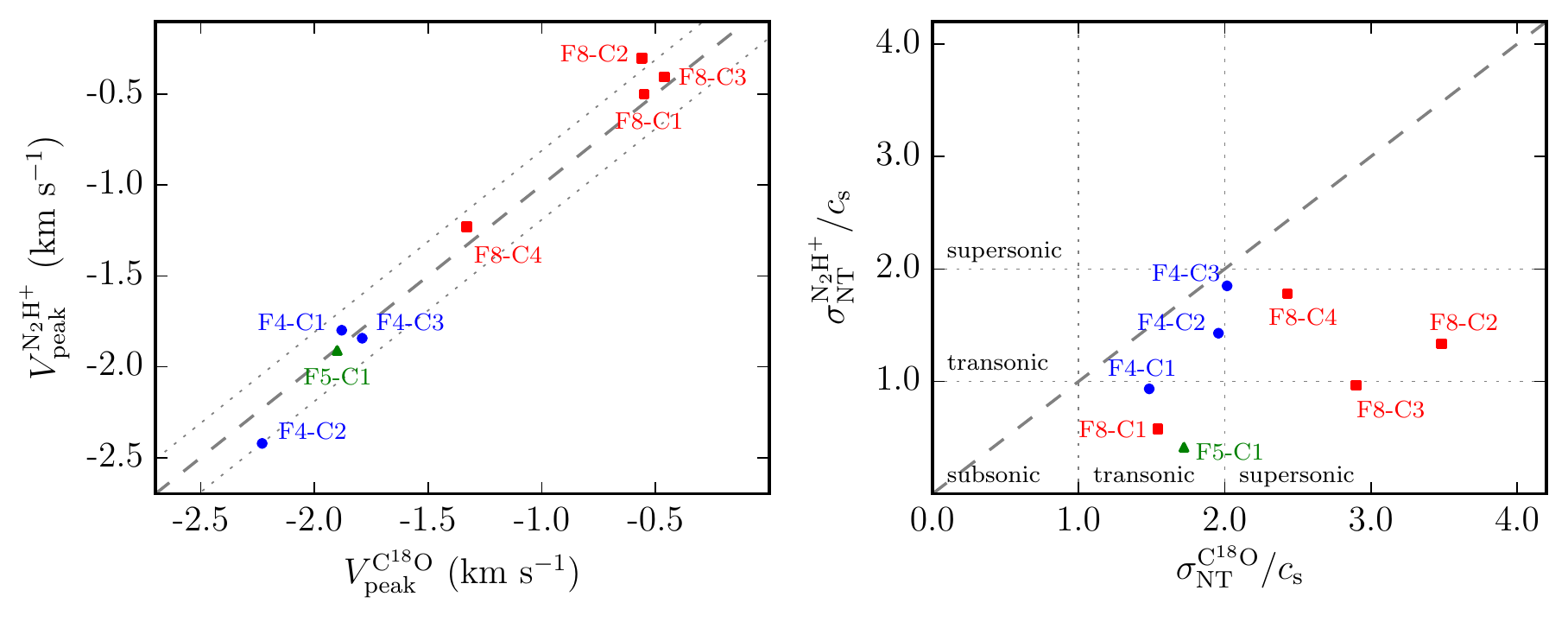} \caption{{\it Left:} Peak velocity of dense cores traced by $\nthp$ and surrounding materials of filaments traced by $\ceo$. The mean error in $\vpeak$ is about 0.02~$\kms$ for both $\nthp$ and $\ceo$, which is comparable or smaller than the point size. The gray dashed and dotted lines indicate the identical line and its displaced line by the sound speed (about 0.19~$\kms$ at 10~K). {\it Right:} Nonthermal velocity dispersion of dense cores traced by $\nthp$ and surrounding materials of filaments traced by $\ceo$. The gray dashed line indicates that $\sigma_{\rm NT}^{\nthp}$ and $\sigma_{\rm NT}^{\ceo}$ are identical. Cores in F4, F5, and F8 are presented with blue circles, green triangles, and red squares, respectively. } \label{fig:vpeakdcs} \end{figure*}

One noticeable thing is that the relationships of $\sigma_{\rm NT}^{\ceo}$ and $\sigma_{\rm NT}^{\nthp}$ are different from filament to filament, i.e., dense cores which are surrounded with materials in supersonic motion in F8 still have subsonic or transonic motions, while velocity dispersion of dense cores in F4 becomes as large as that of their surrounding filament (F4). This different motion between dense cores and surrounding material in filaments can be seen in the histogram of Figure~9, 
too. The nonthermal velocity distribution of $\nthp$ and $\ceo$ in F4 peaks at transonic, but that of $\nthp$ in F8 peaks at transonic while that of $\ceo$ peaks at supersonic. Filament 5 has only one dense core, F5-C1, which appears to be subsonic $\sigma_{\rm NT}^{\nthp} < 0.5 c_{\rm s}$ while the less dense material is transonic $\sigma_{\rm NT}^{\ceo} \sim 1.5 c_{\rm s}$. 

The difference between F4 and F8 implies that the dense cores in F4 and F8 can be formed by different processes. The two filaments are both supercritical and their velocity fields indicate mass flows along the filament in case of F4 and from filaments to hub in case of F8. However, their morphologies and dynamical properties are totally different. F4 has a single, long filament shape, but F8 has a hub-like morphology. The dense cores and surrounding material of F4 are both subsonic or transonic while $\sigma_{\rm NT}^{\ceo}$ of F8 are transonic or supersonic but $\sigma_{\rm NT}^{\nthp}$ are still subsonic or transonic. Hence, dense cores in F8 may be formed by collision of turbulent flows \citep{padoan2001ASPC}, while dense cores in F4 may form with the filaments. In case of F5, its dynamical property is similar to F8, but it has a shape of a long filament like F4. However, if we focus on the region around F5-C1, small filaments can be seen around the bright hub-like structure by F5-C1 (see the bottom left panel in Figure~10). 
The peak velocity distribution shown in Figure~6 
is similar to that of F8, i.e., it has multiple velocity gradients along various directions although its velocity difference is not as large as that of F8. To confirm that F5 and F5-C1 form via the similar mechanism as F8 and its dense cores, more observations with higher spatial resolution are necessary. \\

\section{Summary and Conclusion} \label{sec:summ}

We present dynamical and chemical properties of $\sim$3~pc long filamentary molecular clouds, L1478 in the California MC. We used various molecular lines obtained with the TRAO 14m antenna: $\ceo (1-0)$ as a tracer of filaments and $\nthp (1-0)$ as a dense core tracer. $\rm SO (32-21)$ and $\rm CS (2-1)$ molecular lines are observed to investigate kinematics and chemical evolution of dense cores as well as $\nthp$. We mapped $\sim$1 square degree area in the (1-0) transitions of $\ceo$ and $\tco$ with the noise levels of $\sim$0.1~K[$\rm T_{A}^{\ast}$], and the velocity resolution of 0.1~$\kms$ for both lines. The $\nthp$ (1-0) line observations were made over an area of $\sim$480 square arcminute, where $\ceo$ is bright, with the noise level of $\lesssim 0.06 ~ \rm K[T_{A}^{\ast}]$, and the velocity resolution of $0.06 ~\kms$. SO and CS molecular lines are also observed over $\sim 440$ square arcminute area. The noise level and the velocity resolution are $\lesssim 0.1~ \rm K[T_{A}^{\ast}]$ and $0.06~ \kms$, respectively.

The main results and conclusions are :
\begin{enumerate}
\item From this data, ten filaments are identified with the dendrogram technique. The basic properties of filaments such as length, width, mass, mass per unit length, and mean velocity gradient are derived. We applied the {\sc FellWalker} algorithm to identify dense cores to the $\nthp$ integrated intensity image, and found 8 dense cores in three filaments among the 10 identified filaments.
\item Considering observed mass per unit length and the effective critical mass per unit length for the filaments, we found that three filaments (F4, F5, and F8) among the 10 filaments are supercritical. These three filaments are found to have dense cores, and young stellar objects are also reported to be embedded in F4 and F8. In the supercritical filaments F4, F5 and F8, infall signatures are seen toward the dense cores. From the observational results, we conclude that three supercritical filaments are gravitationally unstable and continuously contracting.
\item Every filament shows coherent velocity field. Multiple and grouped velocity components like fibers are frequently found in filaments. Nonthermal velocity dispersions derived with $\ceo$ and $\nthp$ indicate that dense cores are subsonic or transonic while the surrounding gases are transonic or supersonic. However, the distributions of velocity dispersion are found to be different from filament to filament, i.e., the nonthermal velocity dispersions of dense cores in F4 are similar to those of surrounding material while dense cores in F8 are subsonic or transonic in the transonic or supersonic surrounding filament material. We propose that the formation process of cores and filaments can be different for their morphologies and environments.  \\
\end{enumerate}

Our further forthcoming studies on the physical properties of filaments and dense cores in different star-forming environments will be more helpful to understand how the filaments and dense cores form.

\acknowledgments

We appreciate to the referee and the editor for the valuable comments and suggestions. We thank J. Montillaud for the valuable discussion and comments. This research was supported by Basic Science Research Program through the National Research Foundation of Korea (NRF) funded by the Ministry of Education, Science and Technology (NRF-2016R1A2B4012593). MT acknowledges partial support from grant AYA2016. \\

\software{astrodendro (Rosolowsky 2008), Astropy (http://dx.doi.org/10.1051/0004-6361/201322068), DisPerSE (Sousbie 2011), FellWalker (Berry 2013), FilFinder (Koch \& Rosolowsky 2016; https://github.com/e-koch/FilFinder), GILDAS/CLASS (Pety 2005), PySpecKit (v0.1.20; Ginsburg et al. 2016)}

\bibliography{L1478_190422}


\appendix \label{sec:append}

We used $astrodendro$ Python package to identify filaments with $\ceo$ datacube. The first step of $astrodendro$ algorithm is to make grids of the PPV (Position Position Velocity) space and finds local maxima in every grid. We gave initial parameters of $4 \rm ~pixels ~along~ RA \times 4 ~pixels ~along ~Dec \times 7 ~pixels ~along ~velocity ~axes$ for this step. And then, structures starting from the local maxima merge the surrounding pixels with lower intensities and become large. When the structures meet a neighboring structure, there are two choices. If the difference between the local maxima and the intensity at the meeting point is larger than the critical value (2$\sigma$ is used in this step), the structure is identified as an independent structure. But if the difference is less than 2$\sigma$, then the local maximum point is rejected and merged into the other structure. There is another parameter to assign the structure as an independent structure, the number of associated pixels. For the structures with associated number of pixels less than 5, we discarded the structures. The merging of structures stop when they meet a neighboring structure or a given minimum intensity (1$\sigma$).

Figure~14 
shows the tree diagram of Filaments 6, 7, and 8, and sample spectra with their Gaussian fitting results. Panel (a) shows that F6 consists with three leaves (S1, S2, and S3) and two branches (S4 and S5). The leaf S1 includes the local maximum pixel of F6, merges with S2 which is another leaf from neighboring local maximum, becomes the branch S4, merges again with a leaf S3, and be the branch S5. In panel (b) and (c), the spatial and spectral structures of F6 can be seen. Again, S1 leaf which has the local maximum pixel (colored with pink) is in the middle of S4 (green) and S5 (yellow), S4 (green contour) encloses S1 (pink) and S2 (red), and S5 (yellow) surrounds S3 (orange) and S4 (green). Likewise, in panel (c), the spectral components of S1 (colored with pink) is surrounded by S4 (green) and S5 (yellow). By checking those components of structures, we confirmed that S5 is a single filament with coherent velocity structure and named F6. 

Panel (d) shows the spectra in the blue box of panel (b) which is the junction of F7 and F8. F7 looks to be connected with F8 in the $Herschel$ dust map and the integrated intensity images of $\tco$ and $\ceo$ (see Figure~2). 
However, double peak $\ceo$ spectra can be seen in the two top left boxes of panel (d), and the lower velocity peak components are connected with northern spectra (F8) and the higher velocity peak components are linked with southern spectra (F7). Hence, F7 is supposed to be an independent filament though its northern part seems to merge with F8 gradually. The blue and cyan histograms are the resulted structures from dendrogram and they are identified as an independent filaments. 

The results shown above confirm that dendrogram represents well the hierarchical structures of 3-dimensional data. However, the structures (datacube given by dendrogram) do not include whole spectra of the structure as shown with colored histograms in the panel (c) and (d) of Figure~14. 
Hence, we can not use the datacubes of structures given by dendrogram but carried out gaussian fitting on $\ceo$ spectra to derive the physical properties of filaments such as distribution of peak velocity, velocity dispersion, and mass of filaments (Section~3.3). 
Gaussian fitting is performed automatically with a Python code and the datacube resulted by dendrogram is used for its initial guess. Though the datacube from dendrogram do not fully cover the spectral components, it always contains the peak channel which can be used as a initial guess of central velocity and velocity dispersion for gaussian fitting. Hence, we used the velocity of the channel having maximum intensity at each position and the widths from the datacube resulted by dendrogram algorithm. The fitting results are overlaid on the spectra in Figure~14. 
$\ceo$ spectra with the gaussian fitting results at the overlapped positions of filaments are presented in Figure~15. 
The spectra have clear double peak components and each peaks are assigned to different structures indicating that the filaments identified by dendrogram are distinct structures.

\begin{figure}
\resizebox{1.0\linewidth}{!}{\plotone{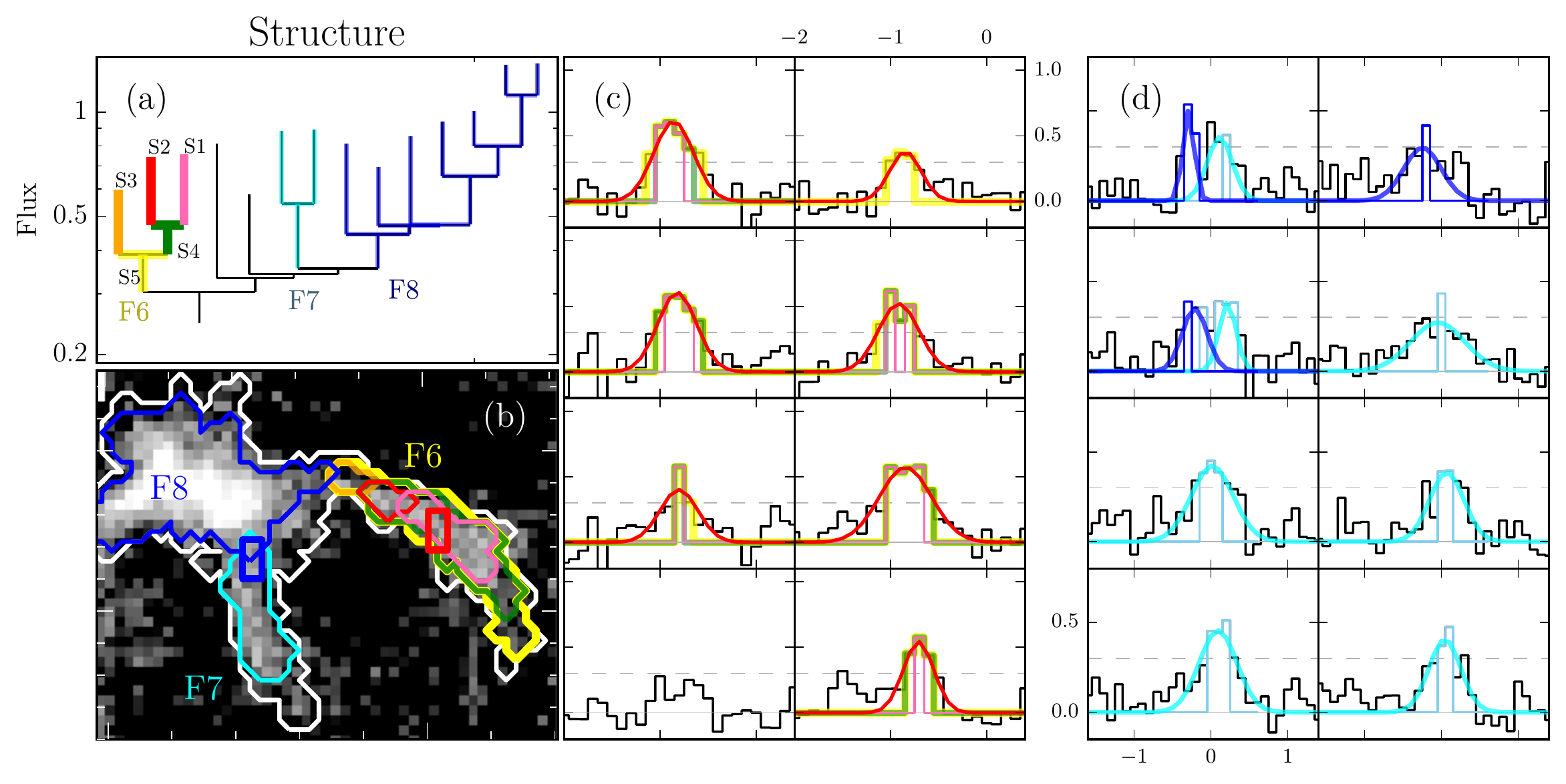}}
\caption{Hierarchical structure of F6, F7, and F8. \textbf{(a)} : Tree diagram of F6, F7, and F8. Leaves (that have no sub-structure, e.g., S1, S2, and S3) and branches (have sub-structures, e.g., S4 and S5) of F6 are presented with different colors. \textbf{(b)} : Contours of leaves and branches derived by dendrogram technique are overlaied on the $\ceo$ moment 0 image to show the spatial distribution of the structures. The same color code with panel (a) is used and the grayscale of integrated intensity map is identical to that of Figure~3. 
\textbf{(c) and (d)} : $\ceo$ spectra of the red and blue box regions of panel (b), respectively. The x- and y- axis are the LSR velocity in $\kms$ and the antenna temperature in kelvin, respectively. The observed spectra are given with black histogram, and those of structures (F7, F8, and sub-structures of F6) are presented with the same color code of panel (a). The gaussian fitting results of filaments are overlaid. The dashed line is 3$\sigma$ level.  \label{fig:F678}} 
\end{figure}

\begin{figure}
\resizebox{1.0\linewidth}{!}{\plotone{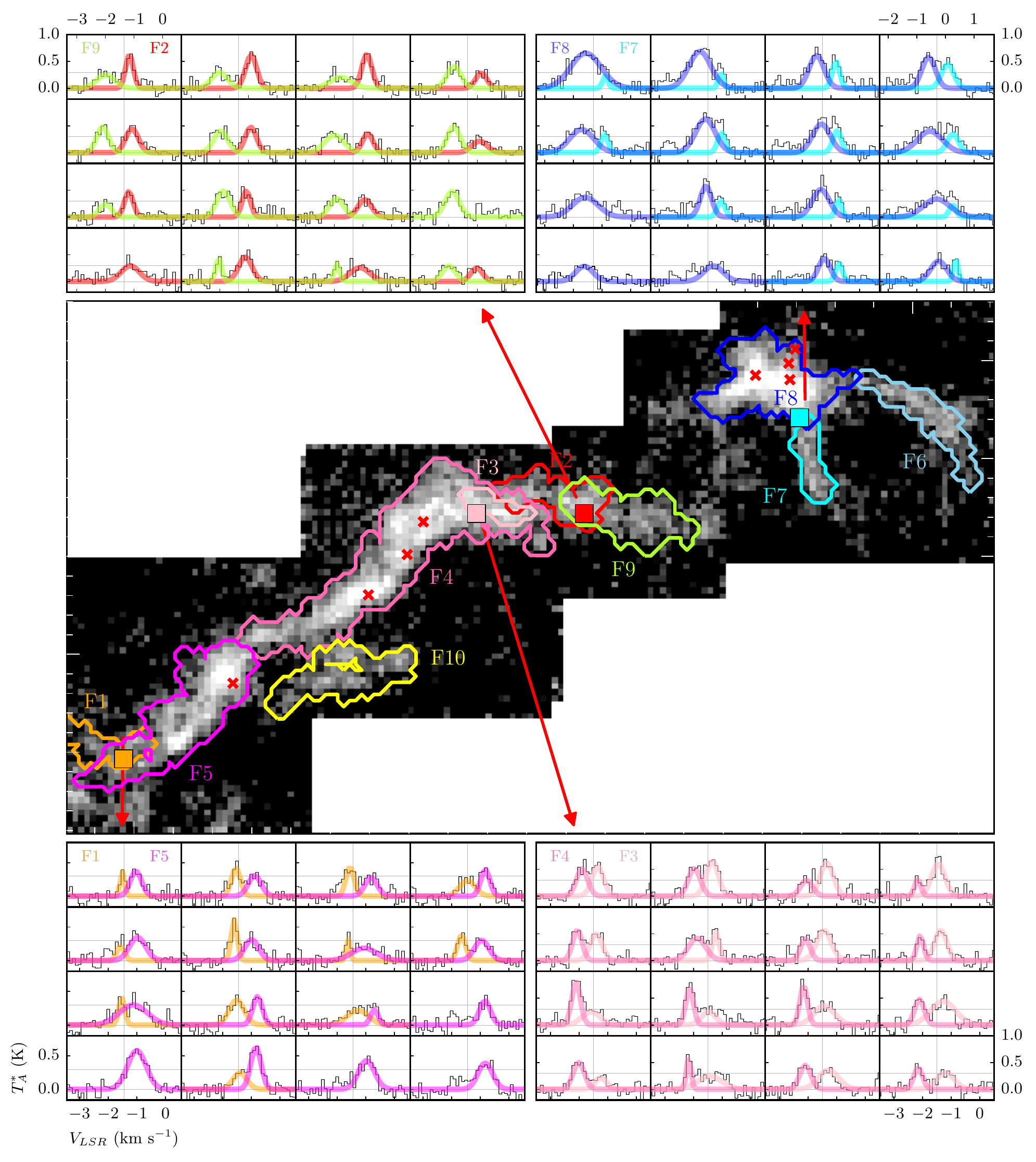}}
\caption{$\ceo$ spectra of filaments and the gaussian fitting results at the areas where filaments are overlapped. (center) : Filaments overlaid on the $\ceo$ integrated intensity map. The squares colored with orange, pink, red, and cyan represent the locations of the spectra shown. (top left) : spectra at the red square of the central image where F2 and F9 meet together. The x- and y- axis of the spectra are the LSR velocity in $\kms$ and the antenna temperature in kelvin, respectively. The gaussian fitting results are plotted with red and green for F2 and F9, respectively. (top right) : The same with top left but for F8 and F7 with blue and cyan colors. (bottom left) : The same with top left but for F1 and F5 with orange and magenta colors. (bottom right) : The same with top left but for F4 and F3 with hotpink and pink colors. \label{fig:gauss}}
\end{figure}

\end{document}